\documentclass[a4paper,12pt,final]{article}
\usepackage{etex}

\usepackage[sc]{mathpazo}

\usepackage[table]{xcolor}

\usepackage{threeparttable}
\usepackage{standalone}
\usepackage{booktabs, dcolumn}
\usepackage{booktabs}

\usepackage{afterpage}

\usepackage{rotating}
\usepackage{tikz}

\usepackage{latexsym}
\usepackage{amsfonts}
\usepackage{amssymb}
\usepackage{amsthm}
\usepackage{amsmath}
\usepackage[mathscr]{eucal}
\usepackage{multicol}
\usepackage{setspace,graphicx}
\usepackage{pdfpages}
\usepackage{pdflscape}
\usepackage{float}
\usepackage{anyfontsize}
\usepackage{t1enc}

\usepackage{caption}
\usepackage{subcaption}
\DeclareCaptionFont{captiofont}{\footnotesize}
\usepackage{verbatim}
\usepackage[plainpages=false,
breaklinks=true,
colorlinks=true,
urlcolor=gray,
citecolor=blue,
linkcolor=red,
bookmarks=true,
bookmarksopen=true,
bookmarksopenlevel=1,
pdfstartview=FitH,
pdfview=FitH]{hyperref}

\usepackage{color,hyperref}

\usepackage[round]{natbib}

\usepackage[lmargin=1.0in, rmargin=1.0in, tmargin=1in, bmargin=1in,paperwidth=8.5in, paperheight=11in]{geometry}

\usepackage{booktabs}
\usepackage{tabulary}
\usepackage{tabularx}
\usepackage{multirow}

\usepackage{enumerate}

\usepackage{sectsty}
\usepackage{lipsum}
\sectionfont{\centering}

\usepackage{tocloft}
\usepackage{titlesec}
\titleformat{\section}
{\large\bfseries}{\thesection}{0.5em}{}

\renewcommand\thesection{\arabic{section}}

\renewcommand\thesubsection{\thesection.\arabic{subsection}}
\titleformat{\subsection}
{\normalfont\itshape}{\thesubsection}{0.5em}{}

\renewcommand\thesubsubsection{\thesection.\arabic{subsection}.\arabic{subsubsection}}
\titleformat{\subsubsection}{\normalfont\itshape}{\thesubsubsection}{0.5em}{}

\usepackage{setspace}

\titlespacing\section{0pt}{10pt plus 4pt minus 2pt}{5pt plus 2pt minus 2pt}
\titlespacing\subsection{0pt}{10pt plus 4pt minus 2pt}{0pt plus 2pt minus 2pt}
\titlespacing\subsubsection{0pt}{10pt plus 4pt minus 2pt}{0pt plus 2pt minus 2pt}

\providecommand{\keywords}[1]{\textbf{Keywords:}  #1}
\providecommand{\JEL}[1]{\textbf{JEL:}  #1}

\makeatletter
\newcommand*{\myfnsymbolsingle}[1]{%
\ensuremath{%
\ifcase#1
\or 
*%
\or 
\dagger
\or 
\ddagger
\or 
\mathsection
\or 
\mathparagraph
\else 
\@ctrerr
\fi
}%
}
\makeatother

\usepackage{alphalph}
\newalphalph{\myfnsymbolmult}[mult]{\myfnsymbolsingle}{}

\renewcommand*{\thefootnote}{%
\myfnsymbolmult{\value{footnote}}%
}

\makeatletter
\def\@xfootnote[#1]{%
\protected@xdef\@thefnmark{#1}%
\@footnotemark\@footnotetext}
\makeatother

\usepackage{bbding} 		

\usepackage[symbol]{footmisc}

\usepackage{chngcntr}

\usepackage{scrwfile}
\TOCclone[\contentsname]{toc}{atoc}

\AfterTOCHead[toc]{%
}
\AfterTOCHead[atoc]{%
\edef\maintocdepth{\the\value{tocdepth}}%
\value{tocdepth}=-10000\relax%
}

\usepackage{blindtext}

\def\RSSevenFive{{\tt SD^{75-25}} }
\def\RSEightZero{{\tt SD^{80-20}} }
\def\RSEightFive{{\tt SD^{85-15}} }
\def\RSNineZero{{\tt SD^{90-10}} }
\def\RSNineFive{{\tt SD^{95-5}} }

\newenvironment{prop*}
  {\ex}
  {\endex}

\newenvironment{remark*}
  {\ex}
  {\endex}

\newenvironment{definition*}
  {\ex}
  {\endex}

\begin{document}

\hypersetup{colorlinks,linkcolor=black} 

\linespread{1}

\title{\Large \setcounter{footnote}{2}Skewness Dispersion and Stock Market Returns\thanks{The support from the Czech Science Foundation under the 24-11555S project is gratefully acknowledged.}
\vspace{20pt}}

\author{\setcounter{footnote}{0}Mykola Babiak\thanks{Lancaster University Management School, Bailrigg, Lancaster, LA1 4YX, United Kingdom, E-mail: \texttt{m.babiak@lancaster.ac.uk}.}
\and
\setcounter{footnote}{6}Jozef Barun\'{i}k\thanks{Institute of Economic Studies, Charles University, Opletalova 26, 110 00, Prague, CR and Institute of Information Theory and Automation, Academy of Sciences of the Czech Republic, Pod Vodarenskou Vezi 4, 18200, Prague, Czech Republic, E-mail: \texttt{barunik@fsv.cuni.cz}.}\\
\and
\setcounter{footnote}{11}Josef Kurka\thanks{Institute of Economic Studies, Charles University, Opletalova 26, 110 00, Prague, CR and Institute of Information Theory and Automation, Academy of Sciences of the Czech Republic, Pod Vodarenskou Vezi 4, 18200, Prague, Czech Republic, E-mail: \texttt{josef.kurka@fsv.cuni.cz}.} \\}

\date{\hspace{2em}}

\maketitle

\begin{abstract}

Cross-sectional dispersion in firm-level realized skewness is significantly and negatively related to future stock market returns. The predictive power of skewness dispersion is robust to in-sample and out-of-sample estimation and is incremental over a broad set of existing predictors, with only a few alternatives retaining independent explanatory ability. Skewness dispersion also delivers substantial economic gains in portfolio allocation. Its forecasting power is concentrated in months with monetary policy announcements, reflecting an information-based mechanism. The empirical evidence suggests that skewness dispersion captures the gradual incorporation of macro news into prices, which is driven by variation in aggregate risk and valuation adjustments.


\vspace{10pt}

\JEL{G11, G12, G14, G17}

\vspace{10pt}

\keywords{Equity risk premium, predictive regression, skewness dispersion, realized skewness}

\end{abstract}

\renewcommand{\thefootnote}{\arabic{footnote}}
\setcounter{footnote}{0}

\newpage

\section{Introduction}
\label{sec:motivation}

Understanding the drivers of the equity risk premium remains a central objective in finance. A large literature has proposed a wide range of macroeconomic and financial predictors of aggregate stock returns. Yet, it is puzzling that, despite a prominent role of higher-order moments, such as skewness, in explaining individual stock returns, their relevance for predicting aggregate market returns remains surprisingly limited. Recent work by \citet*{jondeau2019average} provides the first evidence that the higher-moment information --- the average firm-level skewness --- is informative for future market returns. Focusing on the average, however, masks potentially important variation in skewness across firms. If investors differ in their expectations about asymmetric payoffs, this heterogeneity should be reflected not only in the average level but also in the distribution of skewness across stocks. This observation motivates an examination of whether cross-sectional variation in skewness contains incremental information about future market returns.

In this paper, we show that cross-sectional dispersion in firm-level realized skewness strongly predicts future market returns. Using high-frequency data for a broad sample of 6,770 U.S. stocks from 2000 to 2022, we construct a novel predictor --- skewness dispersion --- defined as the inter-percentile range of the cross-sectional distribution of daily realized skewness computed from intraday returns. This measure captures the extent of heterogeneity in the asymmetry of firm-level return distributions. Periods of high dispersion are characterized by more pronounced positive and negative skewness across stocks, reflecting greater heterogeneity, whereas low dispersion corresponds to a more homogeneous cross-section.

Our analysis demonstrates a significant negative relationship between skewness dispersion and future market returns over 1- to 12-month horizons. The predictive relation is statistically significant under various estimation procedures, including both overlapping and non-overlapping regressions, and inference robust to persistence \citep*{kostakis2015robust}. The results are not driven by recessionary periods and remain strong in out-of-sample tests, where skewness dispersion delivers substantial improvements over the historical average benchmark. Moreover, the predictive content of skewness dispersion is incremental to a comprehensive set of 50 established predictors from the literature.\footnote{Section \ref{sec:controls} presents a detailed description, and Table \ref{tab:variables} in the Appendix provides a concise summary.} Meanwhile, only six other variables exhibit an incremental predictive power in both in-sample and out-of-sample tests.\footnote{These include a skewness factor \citep*{andreou2019information} among skewness-based predictors, none of the stock cross-section variables, three sentiment-related measures --- manager and investor attention indices \citep*{jiang2019manager,chen2022investor}, the aggregate implied volatility spread \citep*{han2021information}, a common volatility risk premium \citep*{babiak2023common}, and the book-to-market ratio across the remaining two groups.} Interestingly, skewness dispersion encompasses both market-level and average firm-level skewness. 

Having established a strong predictive ability of skewness dispersion, we assess the economic value of its forecasts for a mean-variance investor who allocates his wealth between the stock market index and the risk-free bills. We document that the mean-variance investor achieves substantial benefits in terms of certainty equivalent return (CER) and Sharpe ratio relative to the buy-and-hold strategy. For instance, incorporating skewness-dispersion-based forecasts into optimal portfolios yields CER gains of 709 to 875 basis points and Sharpe ratios of 0.82 to 0.91 for a monthly rebalanced portfolio. This compares with 240 basis points and 0.57 for the passive strategy. We observe a similar outperformance for quarterly and semi-annual rebalancing horizons. In general, these are among the highest CER gains and Sharpe ratios compared to the forecasts based on the historical average and alternative predictors. The skewness dispersion forecasts further exhibit a superior performance relative to other benchmarks in terms of cumulative wealth.

We implement several tests to shed light on the underlying economic mechanism. First, we examine both rational and behavioral channels. Skewness dispersion is contemporaneously negatively related to measures of aggregate risk, suggesting that it captures variation in perceived economic conditions. It is also positively associated with several investor sentiment series, though the effects are generally modest and not always significant. Among sentiment proxies, skewness dispersion exhibits the strongest correlations of $0.20$ with investor attention in \cite*{chen2022investor}, and $0.30$ and $-0.28$ with the American Association of Individual Investors (AAII) bullish and bearish measures in the AAII survey data. This result indicates that the behavioral component of skewness dispersion primarily captures the directly measurable sentiment of investor expectations, particularly those underlying the AAII measures. 

Second, we examine the predictability of skewness dispersion conditional on sentiment levels. Motivated by the strong role of investor expectations, we divide our sample into two regimes based on the spread between the bullish and bearish characteristics from the American Association of Individual Investors. We document that the predictive power of skewness dispersion is much stronger when investors are more optimistic, though its predictability remains economically and statistically strong in a pessimistic regime. This supports our behavioral explanation that skewness captures investor expectations; however, a pure behavioral mechanism cannot fully account for the observed predictability.

Third, we observe that, among all predictors studied, cross-sectional variation in firm-level skewness is most closely related to aggregate short selling activity. To the extent that short positions reflect the forward-looking expectations of relatively sophisticated investors, we should expect skewness dispersion to have stronger predictive power during periods of concentrated information arrival, when most investors rapidly incorporate new information into asset prices. We test this conjecture by focusing on monetary policy announcements, which have been shown to be the most important macroeconomic events \citep*{badidi2026macroeconomic}. We show that the predictive power of skewness dispersion is sharply concentrated in FOMC-related months and is absent in non-FOMC periods. Importantly, predictability emerges prior to policy announcements and persists, albeit much more weakly, one month later. 

Overall, this evidence supports a mechanism in which skewness dispersion captures heterogeneity in investor beliefs that is partially incorporated into prices ahead of major information releases and quickly resolved as information is revealed and processed. Such a return predictability, driven by macroeconomic announcements, reflects changes in aggregate risk and the elimination of mispricing, supporting the dual nature of skewness dispersion as both a risk-based and behavioral signal.


Our paper relates to several strands of the literature. First, we contribute to the understanding of how skewness predicts subsequent returns. In the cross-section, a plethora of evidence shows that skewness predicts future individual stock returns and equity option returns.\footnote{See, for example, \cite*{boyer2010expected}, \cite*{boyer2014stock}, \cite*{bali2013does}, \cite*{conrad2013ex}, \cite*{boyer2014stock}, \cite*{amaya2015does}, and \cite*{byun2016gambling}.} In contrast, \cite{jondeau2019average} is, to the best of our knowledge, the only study successfully connecting the average realized skewness across firms to the time-series predictability of stock market returns. We extend this evidence by examining the cross-sectional distribution of skewness and showing that its dispersion contains distinct and economically meaningful information about future market returns.

Second, we contribute to the literature on stock market predictability using variables constructed from diverse datasets.\footnote{Our analysis considers predictors based on the cross-section of stock returns and characteristics \citep*{pollet2010average, kelly2014tail,kelly2013market,maio2016cross,jondeau2019average}, option prices and analyst forecasts \citep*{yu2011disagreement, bakshi2011improving,bekaert2014vix,martin2017expected, andreou2019information,andreou2019dispersion, kilic2019good,han2021information,babiak2023common}, sentiment \citep*{huang2015investor,jiang2019manager,chen2022investor}, and macroeconomic and financial data \citep*{welch2008comprehensive,goyal2024comprehensive}.} Different from previous studies, we employ high-frequency stock returns to construct a novel measure. The skewness dispersion proposed in our study is related to several aggregate risk and expectation-based measures, highlighting its dual nature, yet it is also distinct from the existing variables. Our predictor most strongly correlates with aggregate short interest in \cite*{rapach2016short} and, probably surprisingly, is unrelated to disagreement, dispersion, and tail risk proxies. Compared with existing predictors, our measure is simple to construct, relies solely on return data, and can be computed at higher frequencies. Another distinguishing feature is that we provide direct evidence that the predictive power of skewness dispersion concentrates around major macroeconomic announcements, such as FOMC meetings, highlighting its information-based mechanism.

Third, we contribute to the ongoing debate on stock market predictability. Early evidence by \cite{welch2008comprehensive} documents the limited out-of-sample predictive ability of standard macroeconomic and financial variables, a finding reinforced by \cite*{goyal2024comprehensive} for a broader set of predictors proposed by the recent literature. Consistent with this evidence, we find that most established variables exhibit weak out-of-sample performance in our sample. We contribute to this literature by proposing a novel predictor with robust in-sample and out-of-sample performance, delivering substantial economic value for investors. Moreover, we show that incorporating skewness dispersion enhances the predictive performance of competing predictors.

The remainder of the paper is organized as follows. Section \ref{section: data} describes the data and construction of skewness dispersion. Section \ref{section: quantitative results} presents the empirical results. Section \ref{section: interpretation of the predictive power of skewness dispersion} investigates the underlying economic mechanisms. Section \ref{section: conclusion} concludes.

\section{Data}
\label{section: data}

We collect high-frequency stock prices from Kibot from December 2000 to December 2022. The sample includes 6770 stocks listed on the New York Stock Exchange, the American Stock Exchange, and NASDAQ, which are also available in the Center for Research in Security Prices dataset. We record five-minute prices from 9:30 EST to 16:00 EST and compute five-minute log returns, yielding 78 intraday returns over 6.5 trading hours. The five-minute grid is constructed using the last price recorded within the previous five-minute period. This choice of a 5-minute frequency is common in the literature \citep{amaya2015does} and is motivated by an optimal trade-off between estimator precision and the impact of microstructure noise \citep*{liu2015does}.

\subsection{Constructing the skewness dispersion}

We now construct the daily realized moments of stock returns. Let $p^s_{t, k/K}$  and $r^s_{t, k} = p^s_{t, k / K} - p^s_{t, (k - 1) / K}$ denote the $k$-th intraday log-price and return of a stock $s$ on day $t.$ We first compute a daily realized variance by summing the squares of intraday high-frequency returns:
\begin{equation}
RV^s_t = \sum_{k = 1}^{K} (r^s_{t, k})^2
\end{equation}
We then calculate a daily realized skewness standardized by the realized variance as:
\begin{equation}
\label{eq:realized_skewness}
RS^s_t = \frac{\sqrt{N} \sum\limits_{k = 1}^{K} (r_{t, k}^s)^3}{(RV_t^s)^{3/2}}.
\end{equation}
The realized skewness measures the asymmetry of the daily return's distribution: a negative (positive) value indicates a fatter left (right) tail of the stock's return distribution.\footnote{Note that as sampling frequency increases, the jump contribution and the continuous contribution to cubic variation are separated, and realized skewness in the limit captures mainly the jump part. Consequently, the measure does not capture the leverage effect arising from the correlation between return and variance innovations, and assets with positive jumps on average will have a positive realized skewness and vice versa. This discussion implies that skewness measured from high-frequency data is likely to contain information different from that computed from daily data or options. We refer to \cite{amaya2015does} for a detailed discussion.} 

We aim to examine the relationship between the cross-sectional distribution of realized skewness and stock market returns. Given that we are specifically interested in exploiting the skewness spread in the cross-section, we define skewness dispersion ${\tt SD}^{a - b}_t$ as the difference between the $a$-th and $b$-th percentiles of the skewness estimates on the day $t.$ Since the higher moment estimates might be sensitive to outlying returns, we consider a broader range between the 95th and 5th percentiles, as well as a more conservative inter-quartile range. As our empirical analysis is conducted monthly, we follow \cite{han2021information} and construct the monthly time-series as the average ${\tt SD}^{a - b}$ over the last five trading days of each month. For robustness, we also compute median estimates in the last week of the month.

\begin{figure}[t!]
\centering
\includegraphics[width=1.00\linewidth]{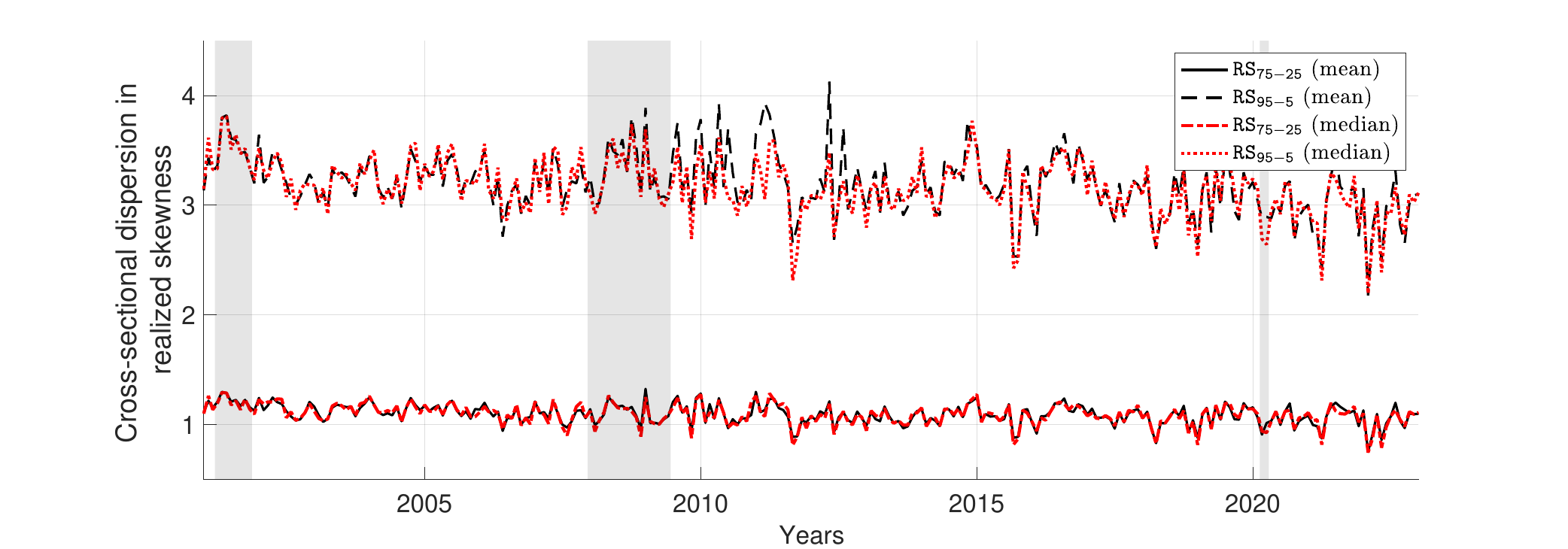}
\caption{Skewness dispersion measures}
\begin{minipage}{\textwidth} \footnotesize This figure illustrates the monthly time-series of skewness dispersion ${\tt SD}^{a - b}_t$ for selected choices of percentiles $a$ and $b.$ The black (red) lines show monthly measures computed as the average (median) ${\tt SD}^{a - b}$ over the last five trading days of each month. The shaded areas denote the NBER recessions. The sample period is from December 2000 to December 2022.
\end{minipage}
\label{figure:dispersion}
\end{figure}

Figure \ref{figure:dispersion} illustrates different specifications of the skewness dispersion. Several observations are noteworthy. First, ${\tt SD}^{95-5}$ tends to be higher and more volatile than ${\tt SD}^{75-25}$ over time. These patterns are expected because the former is more sensitive to outliers and, by construction, measures a wider spread. Second, the specifications based on the average and median values are very similar. Hence, we expect to obtain similar asset pricing predictions in both cases. Third, skewness dispersion varies substantially over time and hovers around 3 (1) for a wider (narrower) inter-percentile range. 

\subsection{Other predictors}
\label{sec:controls}

We employ a comprehensive set of controls to demonstrate that stock market predictability from skewness dispersion is not subsumed by the extant variables proposed in the literature. We collect 50 alternative predictors and divide them into five categories: skewness, stock cross-section, sentiment, variance, and other macroeconomic and financial variables. Table \ref{tab:variables} in the Appendix reports a complete list. 

The first group includes alternative variables associated with skewness. Following \cite{jondeau2019average}, we control for the stock market skewness, the value-weighted average, and the equal-weighted average of monthly skewness values across firms. We retrieve the data from the authors' website. Motivated by a possible link to the option-implied skewness of individual stocks, we use the forward skewness factor recently proposed by \cite{andreou2019information}. The second group consists of those predictors constructed from the cross-section of stocks, such as dispersion in analyst forecasts \citep{andreou2019dispersion,yu2011disagreement} and stock returns \citep*{maio2016cross}, average stock return correlation \citep{pollet2010average}, tail risk estimated from the cross-section of returns \citep*{kelly2014tail}, and the B/M factor \citep{kelly2013market}.

Many recently proposed predictors are based on various sentiment measures. The sentiment-related variables considered in our analysis include manager sentiment \citep{jiang2019manager}, aggregate short interest \citep{rapach2016short}, investor sentiment and attention \citep{huang2015investor,chen2022investor}\footnote{We retrieve the two investor sentiment measures of \cite{huang2015investor} (PLS and PLS$_{\text{orth}}$) from the website of Guofu Zhou. While collecting data from \cite{goyal2024comprehensive}, we noticed an investor sentiment measure (sntm) that is also attributed to \cite{huang2015investor} but differs slightly from the original publication. For completeness, we keep all proxies in our analysis.}, and aggregate implied volatility spread \citep{han2021information}. The fourth category of controls further includes variables related to variance, namely market and average stock-level variances \citep{jondeau2019average}, forward variance factor \citep{andreou2019information}, implied forward variances \citep{bakshi2011improving,martin2017expected}, market and common (bad) variance risk premium \citep{bekaert2014vix,kilic2019good,babiak2023common}.

Finally, the remaining set includes other macroeconomic and financial controls that cannot be clearly categorized into the previous groups. These include fourteen variables in \cite{welch2008comprehensive} and eight variables in \cite{goyal2024comprehensive}. Note that we examine all predictors in the latter study; however, we include some of them in the previous categories.\footnote{We thank all researchers for making their datasets publicly available, which enabled us to collect a comprehensive list of alternative predictors.}

\section{Quantitative results}
\label{section: quantitative results}

\subsection{Univariate predictive regressions}
\label{section: univariate predictive regressions}

We test the ability of skewness dispersion to predict subsequent equity returns by adopting a standard univariate predictive regression:
\begin{equation}
\label{equation: univariate predictive regression}
r_{t, t+h} = \alpha_h + \beta_h {\tt SD}_t^{a-b} + \epsilon_{t, t+h} \quad \text{for} \quad t \in \{1, ..., T - h \},
\end{equation}
where $r_{t, t + h} = \left(1 / h \right) \left( r_{t + 1} + ... + r_{t + h} \right)$, $r_t$ is the S\&P 500 log excess return for month $t,$ and $h$ indicates the predictive horizon of $h$ months. We estimate monthly predictive regressions over horizons of 1, 3, 6, and 12 months. 

We are interested in testing the hypothesis $H_0: \beta_h = 0, H_1: \beta_h \neq 0$. It is well known that statistical inference in Equation (\ref{equation: univariate predictive regression}) is non-trivial due to the \cite*{stambaugh1999predictive} bias and the use of overlapping observations. For this reason, we implement three procedures to tackle this statistical issue. First, we account for the conditional heteroskedasticity and serial correlation using the \cite{newey1986simple} corrected standard errors with $h - 1$ lags. Second, we implement the estimation procedure of \cite{martin2017expected} for non-overlapping observations. Specifically, for each $h$-month horizon, we take the 1st, $(1+h)$-th, $(1+2h)$-th, ... monthly observations of non-overlapping average monthly market excess returns and a predictor, estimate the predictive regression, and save the slope coefficients, t-statistics, and adjusted $R^2.$ Then, we repeat using the 2nd, $(2+h)$-th, $(2+2h)$-th, ... elements. We average each statistic over the $h$ choices of the initial element. Third, we follow the procedure of \cite{kostakis2015robust} and report the IVX-Wald statistics for the slope coefficient. 

\begin{table}[t!]
\centering
\caption{In-sample univariate regression results} 
\begin{minipage}{\textwidth} 
\footnotesize This table reports the in-sample results of univariate predictive regressions. The dependent variable is the average monthly market excess returns in logarithm over the $h$-month horizon. The independent variable is a monthly skewness dispersion measure ${\tt SD}^{a - b},$ calculated as the mean (median) of daily estimates in the last week of the month. Panel A (B) reports the slope coefficients, \cite{newey1986simple} adjusted t-statistics based on $h-1$ lags in brackets, and adjusted $R^2$ of OLS regressions estimated on the sample period from December 2000 to December 2022 (excluding NBER recessions). Panel C reports the regression outputs based on the full sample and non-overlapping observations, following the procedure of \cite{martin2017expected}. The 10\%, 5\%, and 1\% critical values for the t-statistic are 1.645, 1.960, and 2.576. Panel D reports the slope coefficients and IVX-Wald statistics from the \cite{kostakis2015robust} estimation based on the full sample. The 10\%, 5\%, and 1\% critical values for the Wald statistic are 2.706, 3.842, and 6.634. $^{*},^{**},$ and $^{***}$ indicate significance at the 10\%, 5\%, and 1\% levels, respectively.
\medskip
\end{minipage}
\footnotesize
\renewcommand{\arraystretch}{1}
\resizebox{0.95\textwidth}{!}{
\begin{tabular}{lcccccccccc}
\toprule
&       & \multicolumn{4}{c}{Mean}      &       & \multicolumn{4}{c}{Median} \\
\cmidrule{3-6}\cmidrule{8-11}                   
&       & $h = 1$     & $h = 3$     & $h = 6$     & $h = 12$    &       & $h = 1$     & $h = 3$     & $h = 6$     & $h = 12$ \\
\midrule
\multicolumn{11}{l}{Panel A: Full sample} \\
\midrule
$\RSSevenFive$ &       &  -0.099*** &  -0.086*** &  -0.071*** &  -0.050*** &       &  -0.084*** &  -0.073*** &  -0.064*** &  -0.048*** \\
&       & [  -2.95] & [  -4.04] & [  -3.37] & [  -3.42] &       & [  -2.71] & [  -3.64] & [  -3.22] & [  -3.51] \\
$R^2_{\text{adj}}$ &       &    4.09 &    9.19 &   11.18 &    9.50 &       &    3.28 &    7.45 &   10.06 &    9.72 \\
[0.5em]
$\RSEightZero$ &       &  -0.080*** &  -0.068*** &  -0.057*** &  -0.040*** &       &  -0.072*** &  -0.059*** &  -0.051*** &  -0.037*** \\
&       & [  -3.04] & [  -4.03] & [  -3.38] & [  -3.47] &       & [  -2.84] & [  -3.58] & [  -3.16] & [  -3.41] \\
$R^2_{\text{adj}}$ &       &    4.32 &    9.33 &   11.48 &    9.86 &       &    3.77 &    7.36 &   10.15 &    9.29 \\
[0.5em]
$\RSEightFive$ &       &  -0.063*** &  -0.055*** &  -0.045*** &  -0.030*** &       &  -0.055*** &  -0.048*** &  -0.041*** &  -0.029*** \\
&       & [  -3.10] & [  -4.03] & [  -3.32] & [  -3.40] &       & [  -2.69] & [  -3.73] & [  -3.07] & [  -3.32] \\
$R^2_{\text{adj}}$ &       &    4.40 &    9.72 &   11.51 &    9.35 &       &    3.49 &    8.00 &   10.32 &    8.78 \\
[0.5em]
$\RSNineZero$ &       &  -0.050*** &  -0.041*** &  -0.033*** &  -0.022*** &       &  -0.047*** &  -0.041*** &  -0.032*** &  -0.022*** \\
&       & [  -3.17] & [  -3.87] & [  -3.10] & [  -3.14] &       & [  -2.79] & [  -3.78] & [  -2.90] & [  -3.37] \\
$R^2_{\text{adj}}$ &       &    4.64 &    9.47 &   10.61 &    8.18 &       &    4.09 &    9.18 &   10.26 &    8.66 \\
[0.5em]
$\RSNineFive$ &       &  -0.033*** &  -0.028*** &  -0.021*** &  -0.013*** &       &  -0.037*** &  -0.031*** &  -0.024*** &  -0.018*** \\
  &       & [  -3.00] & [  -3.66] & [  -2.76] & [  -2.67] &       & [  -2.86] & [  -3.86] & [  -2.88] & [  -3.39] \\
$R^2_{\text{adj}}$ &       &    4.38 &    9.01 &    9.21 &    6.50 &       &    4.85 &    9.86 &   10.78 &    9.93 \\
\midrule
\multicolumn{11}{l}{Panel B: Excluding NBER recessions} \\
\midrule
$\RSSevenFive$ &       &  -0.095*** &  -0.066*** &  -0.046** &  -0.033*** &       &  -0.077** &  -0.058*** &  -0.042** &  -0.033*** \\
  &       & [  -2.96] & [  -3.23] & [  -2.39] & [  -2.64] &       & [  -2.49] & [  -3.04] & [  -2.25] & [  -2.89] \\
$R^2_{\text{adj}}$ &       &    4.65 &    7.61 &    7.81 &    6.58 &       &    3.48 &    6.56 &    7.15 &    7.40 \\
[0.5em]
$\RSEightZero$ &       &  -0.074*** &  -0.050*** &  -0.037** &  -0.026*** &       &  -0.066*** &  -0.043*** &  -0.032** &  -0.025*** \\
  &       & [  -2.95] & [  -3.12] & [  -2.38] & [  -2.71] &       & [  -2.64] & [  -2.79] & [  -2.17] & [  -2.86] \\
$R^2_{\text{adj}}$ &       &    4.48 &    6.95 &    7.70 &    6.50 &       &    3.88 &    5.49 &    6.50 &    6.41 \\
[0.5em]
$\RSEightFive$ &       &  -0.056*** &  -0.039*** &  -0.028** &  -0.020*** &       &  -0.047** &  -0.033*** &  -0.024** &  -0.018*** \\
  &       & [  -2.90] & [  -3.07] & [  -2.34] & [  -2.70] &       & [  -2.38] & [  -2.82] & [  -2.08] & [  -2.75] \\
$R^2_{\text{adj}}$ &       &    4.23 &    6.70 &    7.48 &    6.11 &       &    3.16 &    5.33 &    5.72 &    5.50 \\
[0.5em]
$\RSNineZero$ &       &  -0.041*** &  -0.027*** &  -0.020** &  -0.014** &       &  -0.038** &  -0.027*** &  -0.018** &  -0.015*** \\
  &       & [  -2.75] & [  -2.90] & [  -2.18] & [  -2.57] &       & [  -2.36] & [  -2.81] & [  -2.00] & [  -2.83] \\
$R^2_{\text{adj}}$ &       &    3.80 &    5.68 &    6.17 &    5.13 &       &    3.29 &    5.54 &    5.27 &    5.98 \\
[0.5em]
$\RSNineFive$ &       &  -0.023** &  -0.016*** &  -0.011* &  -0.008** &       &  -0.026** &  -0.019*** &  -0.013** &  -0.012*** \\
  &       & [  -2.24] & [  -2.73] & [  -1.90] & [  -2.10] &       & [  -2.22] & [  -2.88] & [  -2.01] & [  -2.88] \\
$R^2_{\text{adj}}$ &       &    2.57 &    4.42 &    4.37 &    3.46 &       &    2.96 &    5.35 &    5.39 &    6.70 \\
\bottomrule
\end{tabular}%
}
\label{table: in-sample univariate}%
\end{table}%

\begin{table}[t!]
\centering
\caption*{\footnotesize Table \ref{table: in-sample univariate} \normalfont -- \textit{Continued}}
\footnotesize
\renewcommand{\arraystretch}{1}
\resizebox{1.00\textwidth}{!}{
\begin{tabular}{lcccccccccc}
\toprule
&       & \multicolumn{4}{c}{Mean}      &       & \multicolumn{4}{c}{Median} \\
\cmidrule{3-6}\cmidrule{8-11}                   
&       & $h = 1$     & $h = 3$     & $h = 6$     & $h = 12$    &       & $h = 1$     & $h = 3$     & $h = 6$     & $h = 12$ \\
\midrule
\multicolumn{11}{l}{Panel C: Non-overlapping observations} \\
\midrule
$\RSSevenFive$ &       &  -0.099*** &  -0.083*** &  -0.074** &  -0.052*** &       &  -0.084*** &  -0.071** &  -0.065** &  -0.053*** \\
  &       & [  -2.95] & [  -2.74] & [  -2.55] & [  -2.59] &       & [  -2.71] & [  -2.54] & [  -2.38] & [  -2.67] \\
$R^2_{\text{adj}}$ &       &    4.09 &    9.15 &   12.71 &   13.17 &       &    3.28 &    7.85 &   11.37 &   13.98 \\
[0.5em]
$\RSEightZero$ &       &  -0.080*** &  -0.066*** &  -0.059** &  -0.041*** &       &  -0.072*** &  -0.057** &  -0.051** &  -0.040** \\
  &       & [  -3.04] & [  -2.72] & [  -2.57] & [  -2.62] &       & [  -2.84] & [  -2.42] & [  -2.38] & [  -2.54] \\
$R^2_{\text{adj}}$ &       &    4.32 &    9.32 &   12.88 &   13.42 &       &    3.77 &    7.84 &   11.65 &   13.61 \\
[0.5em]
$\RSEightFive$ &       &  -0.063*** &  -0.053*** &  -0.046** &  -0.031** &       &  -0.055*** &  -0.047** &  -0.041** &  -0.029** \\
  &       & [  -3.10] & [  -2.76] & [  -2.55] & [  -2.53] &       & [  -2.69] & [  -2.55] & [  -2.38] & [  -2.44] \\
$R^2_{\text{adj}}$ &       &    4.40 &    9.80 &   12.72 &   13.21 &       &    3.49 &    8.54 &   11.70 &   12.54 \\
[0.5em]
$\RSNineZero$ &       &  -0.050*** &  -0.040*** &  -0.034** &  -0.023** &       &  -0.047*** &  -0.039** &  -0.033** &  -0.023** \\
  &       & [  -3.17] & [  -2.62] & [  -2.35] & [  -2.32] &       & [  -2.79] & [  -2.52] & [  -2.27] & [  -2.55] \\
$R^2_{\text{adj}}$ &       &    4.64 &    9.65 &   11.90 &   12.31 &       &    4.09 &    9.84 &   11.73 &   12.09 \\
[0.5em]
$\RSNineFive$ &       &  -0.033*** &  -0.027** &  -0.022** &  -0.013* &       &  -0.037*** &  -0.030** &  -0.024** &  -0.018** \\
  &       & [  -3.00] & [  -2.51] & [  -2.08] & [  -1.87] &       & [  -2.86] & [  -2.49] & [  -2.22] & [  -2.67] \\
$R^2_{\text{adj}}$ &       &    4.38 &    9.18 &   10.82 &   11.12 &       &    4.85 &   10.59 &   12.25 &   14.17 \\
\midrule
\multicolumn{11}{l}{Panel D: IVX estimation} \\
\midrule
$\RSSevenFive$ &       &  -0.098*** &  -0.057*** &  -0.039*** &  -0.026*** &       &  -0.083*** &  -0.050*** &  -0.036*** &  -0.026*** \\
IVX-Wald &       & [  10.98] & [  16.54] & [  17.79] & [  13.50] &       & [   8.76] & [  13.62] & [  16.32] & [  13.41] \\
[0.5em]
$\RSEightZero$ &       &  -0.081*** &  -0.045*** &  -0.030*** &  -0.020*** &       &  -0.073*** &  -0.040*** &  -0.028*** &  -0.018*** \\
IVX-Wald &       & [  11.94] & [  16.96] & [  17.75] & [  13.22] &       & [  10.55] & [  13.70] & [  16.01] & [  11.94] \\
[0.5em]
$\RSEightFive$ &       &  -0.064*** &  -0.035*** &  -0.023*** &  -0.014*** &       &  -0.056*** &  -0.033*** &  -0.022*** &  -0.014*** \\
IVX-Wald &       & [  12.31] & [  17.39] & [  17.22] & [  12.27] &       & [   9.79] & [  14.95] & [  16.38] & [  11.31] \\
[0.5em]
$\RSNineZero$ &       &  -0.051*** &  -0.026*** &  -0.016*** &  -0.009*** &       &  -0.047*** &  -0.027*** &  -0.017*** &  -0.010*** \\
IVX-Wald &       & [  13.37] & [  17.07] & [  15.50] & [   9.94] &       & [  11.38] & [  17.13] & [  16.36] & [  11.09] \\
[0.5em]
$\RSNineFive$ &       &  -0.035*** &  -0.017*** &  -0.010*** &  -0.005*** &       &  -0.038*** &  -0.019*** &  -0.011*** &  -0.007*** \\
IVX-Wald &       & [  13.00] & [  16.21] & [  12.99] & [   7.49] &       & [  13.90] & [  17.45] & [  14.78] & [  10.43] \\
\bottomrule
\end{tabular}%
}
\end{table}%

Table \ref{table: in-sample univariate} presents the univariate regression results for five versions of the skewness dispersion measure ${\tt SD}^{a - b}$ depending on a choice of percentiles $a$ and $b$. Panel A reports the slope coefficients, their \cite{newey1986simple} adjusted t-statistics, and the adjusted $R^2$ based on the full sample estimation. All beta coefficients are negative and statistically significant at the 1\% confidence level. The regression outputs remain very similar when the monthly skewness dispersion measure is computed using the median rather than the mean. The coefficients tend to increase in magnitude with the range width used to calculate ${\tt SD}^{a - b}$ due to the level shift in the value of ${\tt SD}^{a - b},$ as shown in Figure \ref{figure:dispersion}. For a particular $h,$ the t-statistic remains stable as we move from ${\tt SD}^{75 - 25}$ to ${\tt SD}^{90 - 10}$ and tends to be the lowest (in absolute terms) for $RS_{95 - 5}.$ Nevertheless, the beta coefficient remains statistically significant at the 1\% confidence level. Overall, the statistical significance across all predictive horizons and predictor specifications clearly demonstrates that skewness dispersion, measured as the cross-sectional skewness range, strongly predicts future stock market returns. 

Panel B shows that, when excluding NBER recessions, the slope coefficients remain statistically significant at least at the 5\% confidence level. As shown in Panel C, estimation based on non-overlapping observations leads to qualitatively and quantitatively similar conclusions. Both panels show that the predictive power of skewness dispersion declines only for specifications with wider skewness ranges ($RS_{90-10}$ to $RS_{95-5}$) and longer predictive horizons, but remains equally strong in other cases. This refutes the concern that predictability based on skewness dispersion is driven by recessionary periods or by overlapping observations used to compute cumulative returns. 

\cite{kostakis2015robust} advocate using the IVX estimation, which differs from standard linear regression by explicitly accounting for the degree of persistence of the regressors. In this case, they find that the predictability of many proposed variables remains strong in the short term but weakens with the predictive horizon, to the point that it almost entirely disappears. We implement the IVX procedure and report the results in Panel D. The slope coefficient is statistically significant at the 1\% confidence level for all predictor specifications and horizons. Intuitively, a significant stock market return predictability by skewness dispersion is robust to controlling for predictor persistence because, as we check in unreported results, the autocorrelation of the cross-sectional skewness range is modest and much lower than that of other predictors in the literature. 

\subsection{Bivariate predictive regressions}
\label{section: bivariate predictive regressions}

We also implement a series of bivariate regressions to test whether the predictive ability of skewness dispersion holds when controlling for other variables established in the literature. Specifically, we estimate a bivariate regression:
\begin{equation*}
r_{t, t+h} = \alpha_h + \beta_{1,h} {\tt SD}_t^{75-25} + \beta_{2,h} x_{t} + \epsilon_{t, t+h} \quad \text{for} \quad t \in \{1, ..., T - h \},
\end{equation*}
where $x_t$ is one of the predictors described in Section \ref{sec:controls}. To save space, we report results for ${\tt SD}^{a - b},$ but those for other skewness dispersion specifications are similar. 

Table \ref{table: in-sample bivariate skewness and cross-section} reports the coefficients, t-statistics, and adjusted R-squared for both regressors in each of the bivariate regressions. Panel A demonstrates that skewness dispersion regains the same level of significance when controlling for each of the four skewness measures. Furthermore, the market skewness and average skewness measures become statistically insignificant when controlling for the cross-sectional skewness range. The forward skewness factor is the only one in this group that remains a significant predictor of future stock market returns for up to half a year. 

\begin{table}[t!]
\centering
\caption{In-sample bivariate regression results} 
\begin{minipage}{\textwidth} 
\footnotesize This table reports the in-sample results of bivariate predictive regressions. The dependent variable is the average monthly market excess returns in logarithm over the $h$-month horizon. The independent variable is the monthly skewness dispersion measure $\RSSevenFive$ (calculated as the mean of daily estimates in the last week of the month) and control variables, one at a time. Panels A, B, C, D, and E present results for five categories of alternative predictors: skewness, stock cross-section, sentiment, variance, and other macroeconomic and financial variable. Each panel reports the slope coefficients, \cite{newey1986simple} adjusted t-statistics based on $h-1$ lags in brackets, and adjusted $R^2$ of OLS regressions estimated on the sample period from December 2000 to December 2022. The 10\%, 5\%, and 1\% critical values for the t-statistic are 1.645, 1.960, and 2.576. $^{*},^{**},$ and $^{***}$ indicate significance at the 10\%, 5\%, and 1\% levels, respectively.
\medskip
\end{minipage}
\footnotesize
\renewcommand{\arraystretch}{1}
\resizebox{1.00\textwidth}{!}{
\begin{tabular}{lcccccclccccc}
\toprule
&       & $h = 1$     & $h = 3$     & $h = 6$     & $h = 12$    &       &       &       & $h = 1$     & $h = 3$     & $h = 6$     & $h = 12$ \\
\midrule
\multicolumn{13}{l}{Panel A: Skewness} \\
\midrule
$\RSSevenFive$ &       &  -0.125*** &  -0.104*** &  -0.091*** &  -0.050*** &       & $\RSSevenFive$ &       &  -0.120*** &  -0.098*** &  -0.088*** &  -0.048*** \\
 &       & [  -3.17] & [  -4.63] & [  -4.25] & [  -2.81] &       &  &       & [  -3.05] & [  -4.41] & [  -4.19] & [  -2.88] \\
Skm   &       &   0.021 &  -0.291 &   0.007 &   0.057 &       & Skew  &       &  -0.100 &  -0.124 &  -0.056 &  -0.034 \\
 &       & [   0.05] & [  -1.04] & [   0.04] & [   0.36] &       &  &       & [  -1.04] & [  -1.59] & [  -1.43] & [  -0.98] \\
$R^2_{\text{adj}}$ &       &    6.34 &   11.90 &   14.97 &    8.64 &       & $R^2_{\text{adj}}$ &       &    6.77 &   13.17 &   15.58 &    9.01 \\
[1.0em]
$\RSSevenFive$ &       &  -0.123*** &  -0.101*** &  -0.090*** &  -0.049*** &       & $\RSSevenFive$ &       &  -0.126*** &  -0.095*** &  -0.095*** &  -0.052*** \\
 &       & [  -3.09] & [  -4.65] & [  -4.29] & [  -2.83] &       &  &       & [  -3.15] & [  -4.12] & [  -4.25] & [  -2.69] \\
Skvw  &       &  -0.075 &  -0.099*** &  -0.036 &  -0.013 &       & FSF   &       &   0.063*** &   0.063*** &   0.032*** &   0.013* \\
 &       & [  -1.05] & [  -2.61] & [  -1.60] & [  -0.72] &       &  &       & [   3.68] & [   7.54] & [   4.81] & [   1.88] \\
$R^2_{\text{adj}}$ &       &    6.95 &   14.26 &   15.61 &    8.73 &       & $R^2_{\text{adj}}$ &       &   13.58 &   27.49 &   24.07 &   11.79 \\
\midrule
\multicolumn{13}{l}{Panel B: Stock cross-section} \\
\midrule
$\RSSevenFive$ &       &  -0.102*** &  -0.089*** &  -0.072*** &  -0.049*** &       & $\RSSevenFive$ &       &  -0.094*** &  -0.076*** &  -0.060*** &  -0.038*** \\
 &       & [  -3.11] & [  -4.41] & [  -3.60] & [  -3.51] &       &  &       & [  -2.84] & [  -3.46] & [  -2.79] & [  -2.92] \\
DISP &       &  -0.254* &  -0.235** &  -0.108 &  -0.044 &       & avgcor &       &   0.012 &   0.025 &   0.028** &   0.027** \\
 &       & [  -1.79] & [  -2.08] & [  -1.19] & [  -0.46] &       &  &       & [   0.44] & [   1.25] & [   2.17] & [   2.57] \\
$R^2_{\text{adj}}$ &       &    5.73 &   13.23 &   12.66 &    9.96 &       & $R^2_{\text{adj}}$ &       &    4.19 &   10.51 &   13.96 &   14.69 \\
[1.0em]
$\RSSevenFive$ &       &  -0.097*** &  -0.085*** &  -0.070*** &  -0.049*** &       & $\RSSevenFive$ &       &  -0.099*** &  -0.086*** &  -0.071*** &  -0.048*** \\
 &       & [  -2.85] & [  -4.03] & [  -3.38] & [  -3.40] &       &  &       & [  -2.96] & [  -4.06] & [  -3.44] & [  -3.47] \\
disag &       &   0.003 &   0.001 &   0.001 &   0.001 &       & tail  &       &   0.070 &   0.021 &   0.063 &   0.089 \\
 &       & [   0.81] & [   0.50] & [   0.43] & [   0.57] &       &  &       & [   0.66] & [   0.26] & [   0.99] & [   1.29] \\
$R^2_{\text{adj}}$ &       &    4.32 &    9.32 &   11.33 &    9.90 &       & $R^2_{\text{adj}}$ &       &    4.25 &    9.24 &   11.89 &   12.13 \\
[1.0em]
$\RSSevenFive$ &       &  -0.100*** &  -0.087*** &  -0.072*** &  -0.049*** &       & $\RSSevenFive$ &       &  -0.099*** &  -0.086*** &  -0.071*** &  -0.049*** \\
 &       & [  -3.03] & [  -4.17] & [  -3.42] & [  -3.38] &       &  &       & [  -3.04] & [  -4.42] & [  -3.66] & [  -3.51] \\
rdsp  &       &  -0.172 &  -0.133 &  -0.063 &   0.021 &       & fbm   &       &  -0.083** &  -0.069** &  -0.048* &  -0.001 \\
 &       & [  -0.69] & [  -0.66] & [  -0.40] & [   0.16] &       &  &       & [  -2.06] & [  -2.34] & [  -1.77] & [  -0.04] \\
$R^2_{\text{adj}}$ &       &    4.36 &    9.67 &   11.37 &    9.54 &       & $R^2_{\text{adj}}$ &       &    6.10 &   13.19 &   14.50 &    9.50 \\
\bottomrule
\end{tabular}%
}
\label{table: in-sample bivariate skewness and cross-section}%
\end{table}%

\begin{table}[t!]
\centering
\caption*{\footnotesize Table \ref{table: in-sample bivariate skewness and cross-section} \normalfont -- \textit{Continued}}
\footnotesize
\renewcommand{\arraystretch}{1}
\resizebox{0.95\textwidth}{!}{
\begin{tabular}{lcccccclccccc}
\toprule
&       & $h = 1$     & $h = 3$     & $h = 6$     & $h = 12$    &       &       &       & $h = 1$     & $h = 3$     & $h = 6$     & $h = 12$ \\
\midrule
\multicolumn{13}{l}{Panel C: Sentiment} \\
\midrule
$\RSSevenFive$ &       &  -0.103** &  -0.079*** &  -0.073*** &  -0.028** &       & $\RSSevenFive$ &       &  -0.088*** &  -0.077*** &  -0.063*** &  -0.041*** \\
 &       & [  -2.56] & [  -3.88] & [  -3.37] & [  -2.58] &       &  &       & [  -2.65] & [  -3.88] & [  -3.12] & [  -3.88] \\
MSI   &       &  -0.007*** &  -0.007*** &  -0.006** &  -0.004** &       & PLS$_\text{orth}$ &       &  -0.008** &  -0.007** &  -0.006** &  -0.004 \\
 &       & [  -2.83] & [  -3.10] & [  -2.59] & [  -2.14] &       &  &       & [  -2.03] & [  -2.39] & [  -2.36] & [  -1.52] \\
$R^2_{\text{adj}}$ &       &    8.82 &   15.83 &   20.01 &   12.45 &       & $R^2_{\text{adj}}$ &       &    6.07 &   12.86 &   16.00 &   13.72 \\
[1.0em]
$\RSSevenFive$ &       &  -0.094*** &  -0.078*** &  -0.061*** &  -0.039** &       & $\RSSevenFive$ &       &  -0.099** &  -0.079*** &  -0.072*** &  -0.034** \\
 &       & [  -2.81] & [  -3.96] & [  -3.70] & [  -2.58] &       &  &       & [  -2.60] & [  -3.91] & [  -4.00] & [  -2.22] \\
SII$_{\text{is}}$ &       &  -0.001 &  -0.002 &  -0.003 &  -0.004 &       & AI &       &  -0.053*** &  -0.050*** &  -0.041*** &  -0.032*** \\
 &       & [  -0.46] & [  -0.95] & [  -1.04] & [  -1.34] &       &  &       & [  -3.01] & [  -3.90] & [  -4.32] & [  -4.16] \\
$R^2_{\text{adj}}$ &       &    4.19 &   10.17 &   13.88 &   17.50 &       & $R^2_{\text{adj}}$ &       &   13.02 &   26.31 &   32.41 &   29.18 \\
[1.0em]
$\RSSevenFive$ &       &  -0.081** &  -0.065*** &  -0.049*** &  -0.029** &       & $\RSSevenFive$ &       &  -0.122*** &  -0.097*** &  -0.091*** &  -0.053*** \\
 &       & [  -2.39] & [  -3.47] & [  -3.14] & [  -2.40] &       &  &       & [  -3.31] & [  -4.81] & [  -4.37] & [  -2.99] \\
SII$_{\text{oos}}$ &       &  -0.004* &  -0.005** &  -0.005** &  -0.005** &       & IVS   &       &   1.046*** &   0.963*** &   0.464*** &   0.065 \\
 &       & [  -1.84] & [  -2.39] & [  -2.30] & [  -2.32] &       &  &       & [   2.91] & [   4.06] & [   3.76] & [   0.79] \\
$R^2_{\text{adj}}$ &       &    5.25 &   14.25 &   20.93 &   26.93 &       & $R^2_{\text{adj}}$ &       &   13.53 &   26.14 &   21.72 &    9.94 \\
[1.0em]
$\RSSevenFive$ &       &  -0.090*** &  -0.077*** &  -0.062*** &  -0.036*** &       & $\RSSevenFive$ &       &  -0.102*** &  -0.092*** &  -0.080*** &  -0.056*** \\
 &       & [  -2.74] & [  -3.95] & [  -3.27] & [  -3.79] &       &  &       & [  -2.96] & [  -4.53] & [  -3.91] & [  -4.03] \\
PLS   &       &  -0.006* &  -0.006** &  -0.006*** &  -0.006*** &       & sntm  &       &  -0.006 &  -0.009 &  -0.011 &  -0.009 \\
 &       & [  -1.68] & [  -2.32] & [  -2.89] & [  -4.05] &       &  &       & [  -0.71] & [  -1.20] & [  -1.28] & [  -0.94] \\
$R^2_{\text{adj}}$ &       &    5.40 &   12.99 &   17.70 &   19.89 &       & $R^2_{\text{adj}}$ &       &    4.25 &    9.87 &   13.88 &   13.37 \\
\midrule
\multicolumn{13}{l}{Panel D: Variance} \\
\midrule
$\RSSevenFive$ &       &  -0.143*** &  -0.117*** &  -0.095*** &  -0.049*** &       & $\RSSevenFive$ &       &  -0.104*** &  -0.088*** &  -0.067*** &  -0.044*** \\
 &       & [  -3.82] & [  -5.29] & [  -4.25] & [  -2.67] &       &  &       & [  -3.12] & [  -3.77] & [  -2.90] & [  -2.66] \\
Vm    &       &  -0.368** &  -0.270*** &  -0.078 &   0.021 &       & rsvix &       &  -0.047 &  -0.021 &   0.034 &   0.054 \\
 &       & [  -2.40] & [  -3.24] & [  -1.20] & [   0.53] &       &  &       & [  -0.37] & [  -0.20] & [   0.58] & [   1.59] \\
$R^2_{\text{adj}}$ &       &   11.51 &   18.70 &   16.00 &    8.73 &       & $R^2_{\text{adj}}$ &       &    4.25 &    9.28 &   11.62 &   11.59 \\
[1.0em]
$\RSSevenFive$ &       &  -0.126*** &  -0.106*** &  -0.092*** &  -0.050*** &       & $\RSSevenFive$ &       &  -0.109*** &  -0.098*** &  -0.082*** &  -0.045*** \\
 &       & [  -3.26] & [  -4.82] & [  -4.32] & [  -2.91] &       &  &       & [  -3.02] & [  -5.08] & [  -4.19] & [  -2.64] \\
Vvw   &       &  -0.685* &  -0.792*** &  -0.366* &  -0.114 &       & vp &       &   0.005 &   0.018* &   0.015** &   0.009 \\
 &       & [  -1.97] & [  -3.28] & [  -1.92] & [  -0.60] &       &  &       & [   0.19] & [   1.92] & [   2.17] & [   1.59] \\
$R^2_{\text{adj}}$ &       &    8.81 &   20.03 &   18.11 &    9.17 &       & $R^2_{\text{adj}}$ &       &    4.96 &   14.99 &   19.09 &   11.28 \\
[1.0em]
$\RSSevenFive$ &       &  -0.122*** &  -0.101*** &  -0.090*** &  -0.049*** &       & $\RSSevenFive$ &       &  -0.113*** &  -0.098*** &  -0.078*** &  -0.040** \\
 &       & [  -3.16] & [  -4.51] & [  -4.26] & [  -2.88] &       &  &       & [  -3.26] & [  -5.06] & [  -3.85] & [  -2.16] \\
Vew   &       &  -0.344 &  -0.371** &  -0.110 &  -0.016 &       & vp$_{\text{bad}}$ &       &  -0.005 &   0.014 &   0.020** &   0.018** \\
 &       & [  -1.57] & [  -2.08] & [  -0.90] & [  -0.15] &       &  &       & [  -0.17] & [   0.83] & [   2.29] & [   2.42] \\
$R^2_{\text{adj}}$ &       &    8.23 &   17.18 &   15.83 &    8.62 &       & $R^2_{\text{adj}}$ &       &    4.94 &   13.27 &   19.17 &   14.02 \\
[1.0em]
$\RSSevenFive$ &       &  -0.155*** &  -0.121*** &  -0.108*** &  -0.056*** &       & $\RSSevenFive$ &       &  -0.129*** &  -0.117*** &  -0.099*** &  -0.059*** \\
 &       & [  -3.47] & [  -4.88] & [  -4.74] & [  -2.84] &       &  &       & [  -3.54] & [  -5.82] & [  -5.17] & [  -4.35] \\
FVF   &       &   0.054** &   0.038** &   0.016 &   0.001 &       & cvp   &       &   0.268** &   0.160*** &   0.155*** &   0.149*** \\
 &       & [   2.30] & [   2.52] & [   1.35] & [   0.07] &       &  &       & [   2.46] & [   2.76] & [   4.31] & [   4.70] \\
$R^2_{\text{adj}}$ &       &   11.58 &   16.93 &   18.87 &    9.62 &       & $R^2_{\text{adj}}$ &       &    9.66 &   17.21 &   24.08 &   23.29 \\
[1.0em]
$\RSSevenFive$ &       &  -0.104*** &  -0.091*** &  -0.072*** &  -0.050*** &       & $\RSSevenFive$ &       &  -0.114*** &  -0.108*** &  -0.091*** &  -0.052*** \\
 &       & [  -3.09] & [  -4.13] & [  -3.26] & [  -3.38] &       &  &       & [  -3.06] & [  -5.57] & [  -4.95] & [  -3.48] \\
impvar &       &  -0.800 &  -0.817 &  -0.229 &  -0.133 &       & cvp$_{\text{bad}}$ &       &   0.157* &   0.097 &   0.132*** &   0.153*** \\
 &       & [  -0.62] & [  -0.76] & [  -0.32] & [  -0.19] &       &  &       & [   1.65] & [   1.10] & [   3.69] & [   4.50] \\
$R^2_{\text{adj}}$ &       &    4.41 &   10.18 &   11.32 &    9.59 &       & $R^2_{\text{adj}}$ &       &    6.54 &   14.22 &   21.95 &   24.37 \\
\bottomrule
\end{tabular}%
}
\end{table}%

\begin{table}[t!]
\centering
\caption*{\footnotesize Table \ref{table: in-sample bivariate skewness and cross-section} \normalfont -- \textit{Continued}}
\footnotesize
\renewcommand{\arraystretch}{1}
\resizebox{0.95\textwidth}{!}{
\begin{tabular}{lcccccclccccc}
\toprule
\multicolumn{13}{l}{Panel E: Other macroeconomic and financial variables} \\
\midrule
$\RSSevenFive$ &       &  -0.088** &  -0.075*** &  -0.058** &  -0.031** &       & $\RSSevenFive$ &       &  -0.092*** &  -0.080*** &  -0.064*** &  -0.042*** \\
 &       & [  -2.55] & [  -3.44] & [  -2.57] & [  -2.47] &       &  &       & [  -2.75] & [  -3.90] & [  -3.19] & [  -3.73] \\
dp    &       &   0.027 &   0.028 &   0.031* &   0.032*** &       & tbl   &       &  -0.325** &  -0.307*** &  -0.316*** &  -0.369*** \\
 &       & [   1.22] & [   1.38] & [   1.90] & [   3.47] &       &  &       & [  -2.10] & [  -2.71] & [  -2.86] & [  -2.78] \\
$R^2_{\text{adj}}$ &       &    5.13 &   12.47 &   18.71 &   23.59 &       & $R^2_{\text{adj}}$ &       &    5.28 &   12.29 &   17.05 &   25.03 \\
[0.2em]
$\RSSevenFive$ &       &  -0.091*** &  -0.078*** &  -0.062*** &  -0.035*** &       & $\RSSevenFive$ &       &  -0.080** &  -0.067*** &  -0.053*** &  -0.032*** \\
 &       & [  -2.71] & [  -4.03] & [  -3.14] & [  -3.18] &       &  &       & [  -2.38] & [  -3.51] & [  -3.19] & [  -2.71] \\
dy    &       &   0.034* &   0.032* &   0.035** &   0.034*** &       & lty   &       &  -0.387* &  -0.392*** &  -0.377** &  -0.367** \\
 &       & [   1.73] & [   1.92] & [   2.55] & [   4.72] &       &  &       & [  -1.92] & [  -2.63] & [  -2.58] & [  -2.25] \\
$R^2_{\text{adj}}$ &       &    5.81 &   13.80 &   20.65 &   25.90 &       & $R^2_{\text{adj}}$ &       &    5.31 &   12.92 &   17.39 &   20.87 \\
[0.2em]
$\RSSevenFive$ &       &  -0.100*** &  -0.089*** &  -0.074*** &  -0.051*** &       & $\RSSevenFive$ &       &  -0.098*** &  -0.086*** &  -0.070*** &  -0.049*** \\
 &       & [  -2.92] & [  -4.09] & [  -3.51] & [  -3.71] &       &  &       & [  -2.93] & [  -3.99] & [  -3.34] & [  -3.39] \\
ep    &       &  -0.001 &  -0.004 &  -0.005 &  -0.002 &       & ltr   &       &   0.056 &  -0.004 &   0.033 &   0.001 \\
 &       & [  -0.11] & [  -0.60] & [  -0.71] & [  -0.41] &       &  &       & [   0.52] & [  -0.06] & [   0.85] & [   0.06] \\
$R^2_{\text{adj}}$ &       &    4.10 &    9.55 &   12.03 &    9.84 &       & $R^2_{\text{adj}}$ &       &    4.25 &    9.19 &   11.48 &    9.50 \\
[0.2em]
$\RSSevenFive$ &       &  -0.100*** &  -0.088*** &  -0.073*** &  -0.050*** &       & $\RSSevenFive$ &       &  -0.100*** &  -0.087*** &  -0.072*** &  -0.053*** \\
 &       & [  -2.98] & [  -4.25] & [  -3.60] & [  -3.48] &       &  &       & [  -2.99] & [  -4.12] & [  -3.63] & [  -3.96] \\
de    &       &   0.005 &   0.007 &   0.008* &   0.006** &       & tms   &       &   0.058 &   0.023 &   0.046 &   0.125 \\
 &       & [   0.60] & [   1.24] & [   1.92] & [   2.47] &       &  &       & [   0.28] & [   0.14] & [   0.31] & [   0.94] \\
$R^2_{\text{adj}}$ &       &    4.30 &   10.65 &   14.59 &   13.07 &       & $R^2_{\text{adj}}$ &       &    4.12 &    9.21 &   11.28 &   10.88 \\
[0.2em]
$\RSSevenFive$ &       &  -0.102*** &  -0.090*** &  -0.071*** &  -0.047*** &       & $\RSSevenFive$ &       &  -0.099*** &  -0.086*** &  -0.071*** &  -0.048*** \\
 &       & [  -3.01] & [  -4.06] & [  -3.23] & [  -3.13] &       &  &       & [  -2.96] & [  -4.02] & [  -3.36] & [  -3.35] \\
svar  &       &  -0.306 &  -0.349 &   0.008 &   0.252** &       & dfy   &       &  -0.419 &  -0.074 &   0.412 &   0.646** \\
 &       & [  -0.33] & [  -0.54] & [   0.03] & [   2.16] &       &  &       & [  -0.44] & [  -0.08] & [   0.65] & [   2.35] \\
$R^2_{\text{adj}}$ &       &    4.30 &    9.99 &   11.18 &   10.95 &       & $R^2_{\text{adj}}$ &       &    4.24 &    9.21 &   11.95 &   13.18 \\
[0.2em]
$\RSSevenFive$ &       &  -0.087*** &  -0.074*** &  -0.057*** &  -0.031*** &       & $\RSSevenFive$ &       &  -0.110*** &  -0.094*** &  -0.080*** &  -0.056*** \\
 &       & [  -2.64] & [  -3.88] & [  -3.13] & [  -2.69] &       &  &       & [  -3.18] & [  -4.10] & [  -3.34] & [  -3.77] \\
bm    &       &   0.096** &   0.103*** &   0.114*** &   0.105*** &       & dfr   &       &   0.205 &   0.147 &   0.158 &   0.099* \\
 &       & [   2.07] & [   3.13] & [   4.26] & [   4.59] &       &  &       & [   0.80] & [   1.09] & [   1.49] & [   1.83] \\
$R^2_{\text{adj}}$ &       &    5.98 &   15.61 &   25.24 &   31.27 &       & $R^2_{\text{adj}}$ &       &    4.86 &   10.35 &   13.59 &   11.29 \\
[0.2em]
$\RSSevenFive$ &       &  -0.102*** &  -0.089*** &  -0.074*** &  -0.052*** &       & $\RSSevenFive$ &       &  -0.099*** &  -0.086*** &  -0.072*** &  -0.049*** \\
 &       & [  -2.99] & [  -4.01] & [  -3.28] & [  -3.62] &       &  &       & [  -2.95] & [  -4.10] & [  -3.65] & [  -3.40] \\
ntis  &       &   0.204 &   0.219 &   0.196 &   0.133 &       & infl  &       &   0.105 &  -0.084 &  -0.702 &  -0.733*** \\
 &       & [   1.06] & [   1.15] & [   0.98] & [   0.79] &       &  &       & [   0.14] & [  -0.11] & [  -1.49] & [  -2.60] \\
$R^2_{\text{adj}}$ &       &    4.73 &   11.36 &   14.33 &   12.30 &       & $R^2_{\text{adj}}$ &       &    4.10 &    9.21 &   13.22 &   13.36 \\
$\RSSevenFive$ &       &  -0.085** &  -0.070*** &  -0.055*** &  -0.034*** &       & $\RSSevenFive$ &       &  -0.099*** &  -0.086*** &  -0.072*** &  -0.050*** \\
 &       & [  -2.55] & [  -3.62] & [  -3.08] & [  -2.97] &       &  &       & [  -2.97] & [  -4.15] & [  -3.44] & [  -3.39] \\
ogap  &       &  -0.089** &  -0.103*** &  -0.103*** &  -0.108*** &       & dtoat &       &  -0.032 &  -0.029 &  -0.029 &  -0.028** \\
 &       & [  -2.22] & [  -3.52] & [  -3.45] & [  -2.86] &       &  &       & [  -1.01] & [  -1.08] & [  -1.42] & [  -2.39] \\
$R^2_{\text{adj}}$ &       &    5.52 &   14.96 &   21.56 &   31.24 &       & $R^2_{\text{adj}}$ &       &    4.67 &   10.52 &   13.67 &   14.02 \\
[0.2em]
$\RSSevenFive$ &       &  -0.099*** &  -0.089*** &  -0.076*** &  -0.055*** &       & $\RSSevenFive$ &       &  -0.100*** &  -0.089*** &  -0.071*** &  -0.049*** \\
 &       & [  -2.88] & [  -4.17] & [  -3.61] & [  -3.75] &       &  &       & [  -2.98] & [  -3.93] & [  -3.32] & [  -3.30] \\
ndrbl &       &  -0.006 &  -0.040 &  -0.070 &  -0.081* &       & wtexas &       &   0.007 &   0.028 &   0.003 &  -0.002 \\
 &       & [  -0.08] & [  -0.78] & [  -1.57] & [  -1.76] &       &  &       & [   0.23] & [   1.44] & [   0.29] & [  -0.30] \\
$R^2_{\text{adj}}$ &       &    4.09 &    9.62 &   13.57 &   15.69 &       & $R^2_{\text{adj}}$ &       &    4.12 &   10.65 &   11.22 &    9.54 \\
[0.2em]
$\RSSevenFive$ &       &  -0.101*** &  -0.088*** &  -0.072*** &  -0.050*** &       & $\RSSevenFive$ &       &  -0.100*** &  -0.087*** &  -0.072*** &  -0.053*** \\
 &       & [  -3.01] & [  -4.18] & [  -3.67] & [  -4.02] &       &  &       & [  -2.96] & [  -3.85] & [  -3.12] & [  -3.56] \\
tchi  &       &   0.003 &   0.003 &   0.003 &   0.002 &       & lzrt  &       &   0.005 &   0.004 &   0.003 &   0.006 \\
 &       & [   1.35] & [   1.59] & [   1.62] & [   1.18] &       &  &       & [   0.44] & [   0.44] & [   0.34] & [   0.63] \\
$R^2_{\text{adj}}$ &       &    5.20 &   12.23 &   16.80 &   12.71 &       & $R^2_{\text{adj}}$ &       &    4.20 &    9.44 &   11.43 &   11.15 \\
[0.2em]
$\RSSevenFive$ &       &  -0.098*** &  -0.086*** &  -0.070*** &  -0.048*** &       & $\RSSevenFive$ &       &  -0.099*** &  -0.089*** &  -0.074*** &  -0.051*** \\
 &       & [  -2.92] & [  -3.84] & [  -3.05] & [  -2.99] &       &  &       & [  -2.91] & [  -4.08] & [  -3.51] & [  -3.72] \\
dtoy  &       &  -0.007 &  -0.002 &  -0.008 &  -0.017 &       & ygap  &       &  -0.000 &  -0.004 &  -0.004 &  -0.002 \\
 &       & [  -0.14] & [  -0.04] & [  -0.21] & [  -0.79] &       &  &       & [  -0.06] & [  -0.52] & [  -0.63] & [  -0.31] \\
$R^2_{\text{adj}}$ &       &    4.11 &    9.20 &   11.29 &   10.46 &       & $R^2_{\text{adj}}$ &       &    4.09 &    9.47 &   11.87 &    9.71 \\
\bottomrule
\end{tabular}%
}
\end{table}%

Panel B shows that none of the predictors in the stock cross-section group subsumes the predictive ability of skewness dispersion. Furthermore, the coefficients associated with $\RSSevenFive$ remain very close to those obtained from the univariate regression. Thus, the information contained in the cross-sectional distribution of realized skewness measures is unrelated to cross-sectional information in analyst forecasts, stock returns, and book-to-market ratios. Only when controlling for the average correlation among daily stock returns, the slope coefficient of a skewness dispersion variable becomes slightly smaller in absolute terms. However, it remains statistically significant at the 1\%  level. 

Panel C reports the outputs for bivariate regressions controlling for investor sentiment measures. Several variables remain significant predictors of future market returns. For instance, the slope coefficients for the manager and investor sentiment, attention, and aggregate short interest indices, and the implied volatility spread tend to be significant at the 5\% confidence level. Yet, skewness dispersion remains a strong predictor in all cases. Controlling for variance-related variables, Panel D shows that only the common volatility risk premium has both short- and long-term predictive power, which, however, does not subsume the $\RSSevenFive$ predictability. Finally, Panel E shows that including a wide range of other macroeconomic and financial variables in the regression has a minimal effect on the predictive power of skewness dispersion. In turn, $\RSSevenFive$ makes a vast majority of characteristics insignificant, except for dividend yield, book-to-market ratio, output gap, T-bill, and long-term yields. 

In another exercise, we perform bivariate estimation based on the method of \cite{kostakis2015robust}. For brevity, we discuss the key findings and report the detailed results in  Table \ref{table: in-sample bivariate skewness and cross-section ivx} in the appendix. First, we show that the strong predictive power of skewness dispersion is robust to the IVX estimation even in the bivariate case. Second, predictability from alternative variables generally weakens, either for selected horizons or for all horizons. In particular, the forward skewness factor, a single B/M factor, the manager and attention indices, the implied volatility spread, and the common volatility risk premium remain significant at conventional levels. However, some variables that appeared strongly significant in Table \ref{table: in-sample bivariate skewness and cross-section} lose significance altogether. These include (1) DISP from the stock-cross-section group; (2) SII$_{\text{is}}$, PLS, and PLS$_{\text{orth}}$ from the sentiment category; (3) dy, ogap, tbl, and lty among other macroeconomic and financial variables. 

Overall, the bivariate regressions using standard OLS or IVX estimation demonstrate that the relationship between cross-sectional skewness dispersion and subsequent stock market returns remains robust after controlling for numerous previously identified predictors. A small number of existing predictors appear to be significant in the presence of skewness dispersion, especially once the variable persistence is taken into account. 

\subsection{Out-of-sample tests}

This section conducts out-of-sample tests to address the concern that in-sample predictability might not translate into strong out-of-sample performance. For all out-of-sample results, the estimation is based on an expanding window from December 2005 $(t_0)$ to December 2022 $(T)$.

\subsubsection{Out-of-sample $R^2$}

We begin by implementing a univariate regression with a skewness dispersion measure as a predictor. For each month $m$ in the out-of-sample period $[t_0, t_0+1, ..., T-1,  T],$ we estimate a predictive regression:
\begin{equation*}
\text{Model $L$:} \quad r_{t-h, t} = \alpha_h + \beta_{h} {\tt SD}_{t-h}^{a-b} + \epsilon_{t-h, t} \quad \text{for} \quad t \in \{1, ..., m\},
\end{equation*}
where $r_{t-h, t} = \left(1 / h \right) \left( r_{t - h + 1} + ... + r_{t} \right),$ $r_t$ is the S\&P 500 log excess return for month $t,$ and $h$ indicates the predictive horizon of $h$ months. Let $\hat{r}_{m, m+h}^L \equiv \alpha_h + \beta_{h} {\tt SD}_{m}^{a-b}$ denote the fitted value based on the estimated coefficients and ${\tt SD}_{m}^{a-b}$. We compare this forecast of Model $L$ with the benchmark forecast of Model $B,$ which uses the historical average $\hat{r}_{m, m+h}^B \equiv \frac{1}{m-h}\sum\limits_{t=1}^{m-h} r_{t, t+h}.$ 

Having repeated this procedure for each month in the out-of-sample period, we calculate the out-of-sample $R^2$ as follows:
\begin{equation}
R_{OOS}^2 = 1 - \frac{MSFE_L}{MSFE_B},
\label{eq:oos_R2}
\end{equation}
where $MSFE_L$ is the mean squared forecast error for a predictive regression, and $MSFE_B$ is the mean squared forecast error for a benchmark forecast equal to the historical average. Following \cite{clark2007approximately}, we further test the hypothesis $H_0: MSFE_B \le MSFE_L, H_1: MSFE_B \geq MSFE_L$ (or equivalently $H_0: R_{OOS}^2 \le 0, H_1: R_{OOS}^2 \geq 0$). Specifically, we compute the difference between two MSFEs, adjusted for the noise associated with the larger model's forecast:
\begin{equation}
\hat{f}_{t,t+h} = \left(r_{t,t+h} - \hat{r}_{t, t+h}^B\right)^2 - \left[\left(r_{t,t+h} - \hat{r}_{t, t+h}^L\right)^2 - \left(\hat{r}_{t, t+h}^B - \hat{r}_{t, t+h}^L\right)^2\right].
\end{equation}
We calculate the \cite{newey1986simple} adjusted t-statistics with $h-1$ lags to test the null hypothesis.

\begin{table}[!t]
\centering
\caption{\footnotesize Out-of-sample univariate regression results}
\begin{minipage}{\textwidth} 
\footnotesize This table reports the out-of-sample results of univariate predictive regressions. The out-of-sample estimation is based on an expanding window from December 2005 to December 2022. The dependent variable is the average monthly market excess returns in logarithm over the $h$-month horizon. The independent variable is a monthly skewness dispersion measure ${\tt SD}^{a - b},$ calculated as the mean (median) value of daily estimates in the last week of the month. The table reports the out-of-sample $R^2$ (in percentages) computed as $R^2_{oos} = 1 - MSFE_L/MSFE_B,$ where $MSFE_B$ is the mean squared forecast error for a benchmark forecast equal to the historical average and $MSFE_L$ is the mean squared forecast error for a predictive regression using a particular predictor. We test the hypothesis $H_0: MSFE_B \le MSFE_L, H_1: MSFE_B \geq MSFE_L$ (or equivalently $H_0: R_{OOS}^2 \le 0, H_1: R_{OOS}^2 \geq 0$) by implementing the \cite{clark2007approximately} test and reporting the resulting \cite{newey1986simple} adjusted t-statistic based on $h-1$ lags in brackets. The 10\%, 5\%, and 1\% critical values for the t-statistic (a one-sided test) are 1.282, 1.645, and 2.334. $^{*},^{**},$ and $^{***}$ indicate significance at the 10\%, 5\%, and 1\% levels, respectively.
\medskip
\end{minipage}
\footnotesize
\begin{tabular}{lcccccccc}
\toprule
  &       & \multicolumn{3}{c}{Mean}      &      &       \multicolumn{3}{c}{Median} \\
\cmidrule{3-5}\cmidrule{7-9}          
&       & $h = 1$     & $h = 3$     & $h = 6$     &       & $h = 1$     & $h = 3$     & $h = 6$     \\
\midrule
$\RSSevenFive$ &       &    2.91*** &    6.12*** &    7.85*** &          &    2.31** &    4.74*** &    6.56*** \\
 &       & [   2.49] & [   3.95] & [   3.71] &        & [   2.20] & [   3.47] & [   3.68] \\
[1.0em]
$\RSEightZero$ &       &    3.15*** &    6.48*** &    8.33*** &        &    2.67*** &    5.34*** &    7.15*** \\
 &       & [   2.59] & [   4.07] & [   3.83] &       & [   2.36] & [   3.65] & [   3.69] \\
[1.0em]
$\RSEightFive$ &       &    3.24*** &    6.68*** &    8.20*** &        &    2.58** &    6.13*** &    6.87*** \\
 &       & [   2.68] & [   4.20] & [   3.83] &      & [   2.26] & [   3.93] & [   3.62] \\
[1.0em]
$\RSNineZero$ &       &    3.45*** &    6.38*** &    7.23*** &         &    3.17*** &    7.49*** &    6.92*** \\
 &       & [   2.79] & [   4.14] & [   3.72] &       & [   2.44] & [   4.22] & [   3.47] \\
[1.0em]
$\RSNineFive$ &       &    3.10*** &    5.69*** &    4.30*** &         &    3.97*** &    8.90*** &    7.57*** \\
 &       & [   2.58] & [   3.66] & [   3.15] &       & [   2.63] & [   4.34] & [   3.44] \\
\bottomrule
\end{tabular}%
\label{table: out-of-sample univariate}%
\end{table}%

Table \ref{table: out-of-sample univariate} shows the results for different definitions of skewness dispersion and predictive horizons. $R_{OOS}^2$ ranges from 2.31\% to 3.97\% for the 1-month horizon, from 4.74\% to 8.90\% for 1-quarter forecast, and from 4.30\% to 8.33\% for the 6-month horizon. All but one of the t-statistics are significant at the 1\% level, indicating that the forecasting regressions using skewness dispersion provide significantly better out-of-sample forecasts than the historical average. This is an important result, as many predictors proposed in the literature fail to outperform the historical average in out-of-sample forecasting \citep{welch2008comprehensive,goyal2024comprehensive}. Interestingly, the monthly skewness dispersion measure based on median daily values tends to yield higher out-of-sample $R^2$ than the one based on average daily values. Intuitively, the median is less sensitive to extreme outliers than the average, providing a more accurate estimate of skewness dispersion.

\begin{figure}[t!]
\centering
\includegraphics[width=1.00\linewidth]{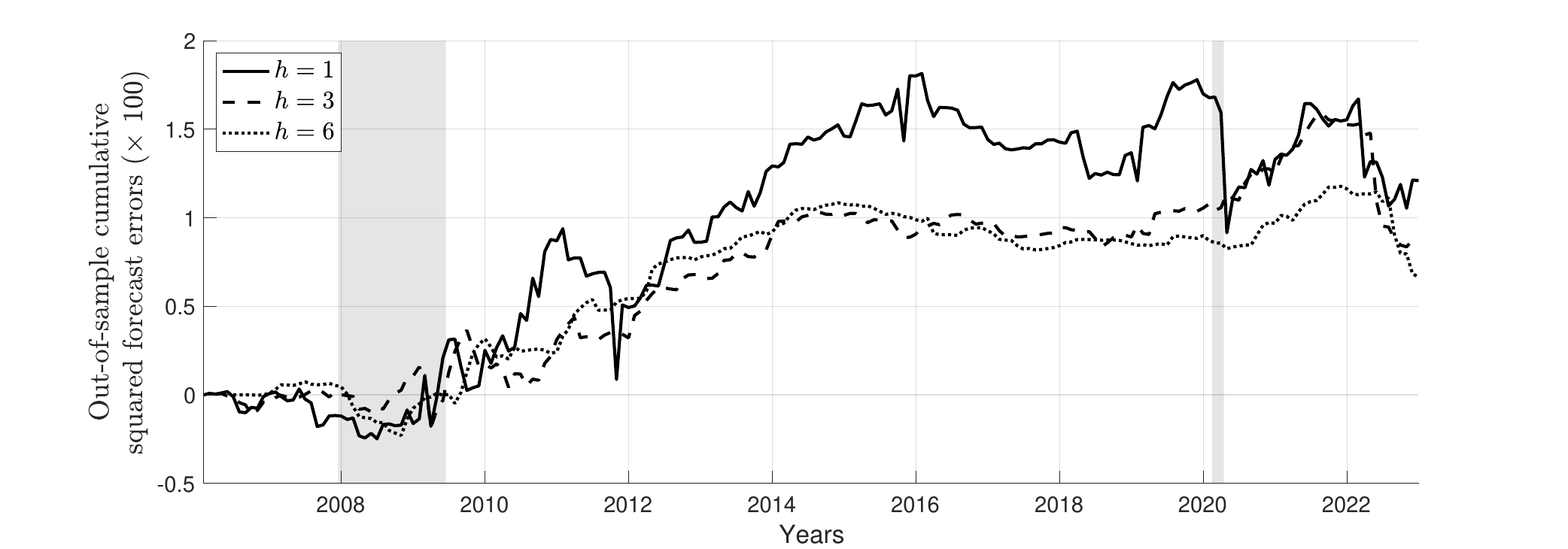}
\caption{Out-of-sample cumulative squared forecast errors}
\begin{minipage}{\textwidth} \footnotesize This figure shows the cumulative difference in out-of-sample squared forecast errors between the benchmark forecast based on the historical average and a univariate predictive regression using $\RSSevenFive,$ calculated as the mean of daily estimates in the last week of the month. The out-of-sample estimation is based on an expanding window from December 2005 to December 2022. We consider forecast errors for $h$-month ahead forecast horizons. The shaded bars indicate NBER recession periods.
\end{minipage}
\label{figure: out of sample cumulative squared forecast errors}
\end{figure}

To better understand the out-of-sample forecast performance over time, Figure \ref{figure: out of sample cumulative squared forecast errors} shows the cumulative difference in squared forecast errors of the benchmark model based on the historical average and the univariate predictive regression using $\RSSevenFive,$ which is measured as the mean of daily values. A positive difference indicates that the benchmark model has larger forecast errors; that is, the skewness dispersion model performs better. The forecast difference at the 1-month horizon shows the largest fluctuations, with pronounced declines at the end of 2011, during the COVID-19 recession, and in 2022. The performance differences at the 3-month and 6-month horizons grow steadily over time and decline slightly in 2022. Overall, this figure demonstrates that skewness dispersion outperforms the historical average across the entire out-of-sample period, and that selected events do not drive its superior performance.  

Next, we implement the out-of-sample bivariate regression with skewness dispersion and alternative controls, one at a time. For each month $m$ in the out-of-sample period $[t_0, t_0+1, ..., T-1,  T],$ we estimate a predictive regression:
\begin{equation*}
\text{Model $L$:} \quad r_{t-h, t} = \alpha_h + \beta_{1,h} {\tt SD}_{t-h}^{75-25} + \beta_{2,h} x_{t-h} + \epsilon_{t-h, t} \quad \text{for} \quad t \in \{1, ..., m\},
\end{equation*}
where $x_t$ is one of other predictors. We denote the fitted value by $\hat{r}_{m, m+h}^L \equiv \alpha_h + \beta_{1,h} {\tt SD}_{m}^{75-25} + \beta_{2,h} x_m.$ We compare the predictive accuracy of this model with several nested benchmarks. We consider the first forecast equal to the historical average. Next, we implement the univariate regression using alternative predictors and use the resulting fitted values as benchmark forecasts. Finally, the third benchmark is the univariate regression with skewness dispersion as a predictor. 

\begin{table}[t!]
\centering
\caption{Out-of-sample bivariate regression results}
\begin{minipage}{1.0\textwidth} 
\footnotesize This table reports the out-of-sample results of univariate and bivariate predictive regressions. The out-of-sample estimation is based on an expanding window from December 2005 to December 2022. The dependent variable is the average monthly market excess returns in logarithm over the $h$-month horizon. The second through fourth columns correspond to the univariate regressions, in which the independent variable is the monthly skewness dispersion measure $\RSSevenFive$ (calculated as the mean of daily estimates in the last week of the month) and control variables, one at a time. The fifth through last columns correspond to the bivariate regressions, in which the independent variables are the skewness dispersion measure $\RSSevenFive$ combined with control variables, one at a time. The table reports the out-of-sample $R^2$ (in percentages) computed as $R^2_{oos} = 1 - MSFE_L/MSFE_B,$ where $MSFE_B$ is the mean squared forecast error for a benchmark forecast and $MSFE_L$ is the mean squared forecast error for a predictive regression using particular predictors. For all univariate predictive regressions, we use the historical average as a benchmark forecast. For the bivariate regressions, we employ the historical average, a univariate predictive regression by an alternative predictor (not $\RSSevenFive$), and a univariate predictive regression with $\RSSevenFive$ as the benchmark forecasts. We test the hypothesis $H_0: MSFE_B \le MSFE_L, H_1: MSFE_B \geq MSFE_L$ (or equivalently $H_0: R_{OOS}^2 \le 0, H_1: R_{OOS}^2 \geq 0$) by implementing the \cite{clark2007approximately} test and reporting the resulting \cite{newey1986simple} adjusted t-statistic based on $h-1$ lags in brackets. The 10\%, 5\%, and 1\% critical values for the t-statistic (a one-sided test) are 1.282, 1.645, and 2.334. $^{*},^{**},$ and $^{***}$ indicate significance at the 10\%, 5\%, and 1\% levels, respectively.
\medskip
\end{minipage}
\footnotesize
\renewcommand{\arraystretch}{1}
\resizebox{1.0\textwidth}{!}{  
\begin{tabular}{lccccccccccccccccc}
\toprule
&       & \multicolumn{3}{c}{Historical average (univariate)}       &       & \multicolumn{3}{c}{Historical average (bivariate)}       &       & \multicolumn{3}{c}{Alternative predictor}       &       & \multicolumn{3}{c}{$\RSSevenFive$}  \\
\cmidrule{3-5}\cmidrule{7-9}\cmidrule{11-13}\cmidrule{15-17}          &       & $h = 1$     & $h = 3$     & $h = 6$     &       & $h = 1$     & $h = 3$     & $h = 6$     &       & $h = 1$     & $h = 3$     & $h = 6$     &       & $h = 1$     & $h = 3$     & $h = 6$ \\
\midrule
$\RSSevenFive$ &       &    2.91*** &    6.12*** &    7.85*** &       &       &       &       &       &       &       &       &       &       &       &  \\
\midrule
\multicolumn{17}{l}{Panel A: Skewness} \\
\midrule
Skm &       &   -3.04 &   -2.65 &   -1.81 &       &    2.56** &    6.31*** &   11.65*** &       &    5.43*** &    8.72*** &   13.22*** &       &   -3.60 &   -3.77 &   -3.25 \\
[0.5em]
Skvw  &       &   -5.89 &    2.88** &    1.14** &       &   -0.40 &   11.67*** &   14.75*** &       &    5.19*** &    9.06*** &   13.76*** &       &   -6.74 &    2.18** &    0.37 \\
[0.5em]
Skew  &       &   -0.56 &    0.53** &    0.27* &       &    4.93*** &    9.33*** &   13.92*** &       &    5.46*** &    8.85*** &   13.69*** &       &   -1.07 &   -0.41 &   -0.60 \\
[0.5em]
FSF   &       &    5.51* &   24.19** &   10.78** &       &   10.86** &   30.80*** &   22.95*** &       &    5.66*** &    8.72*** &   13.65*** &       &    4.01* &   22.83** &    8.00* \\
\midrule
\multicolumn{17}{l}{Panel B: Stock cross-section} \\
\midrule
DISP  &       &   -0.43 &   -1.06 &   -6.46 &       &    3.25*** &    7.00*** &    4.16*** &       &    3.66*** &    7.98*** &    9.98*** &       &    0.35* &    0.95 &   -4.01 \\
[0.5em]
disag &       &   -4.82 &  -16.41 &  -39.43 &       &   -1.10 &   -6.78 &  -24.27 &       &    3.55*** &    8.27*** &   10.87*** &       &   -4.13 &  -13.73 &  -34.86 \\
[0.5em]
rdsp  &       &   -2.74 &  -10.12 &  -11.80 &       &    0.83** &   -1.69 &   -0.92 &       &    3.48*** &    7.65*** &    9.73*** &       &   -2.14 &   -8.32 &   -9.53 \\
[0.5em]
avgcor &       &   -2.64 &   -4.15 &   -2.99 &       &    0.36** &    1.54*** &    3.00*** &       &    2.93*** &    5.46*** &    5.82*** &       &   -2.62 &   -4.87 &   -5.27 \\
[0.5em]
tail  &       &   -0.66 &   -1.50 &   -1.87 &       &    2.36*** &    4.81*** &    6.86*** &       &    3.00*** &    6.22*** &    8.57*** &       &   -0.56 &   -1.39 &   -1.07 \\
[0.5em]
fbm   &       &    1.28* &    2.93** &    1.41 &       &    4.16*** &    8.36*** &    8.79*** &       &    2.91*** &    5.60*** &    7.48*** &       &    1.28* &    2.39* &    1.01 \\
\midrule
\multicolumn{17}{l}{Panel C: Sentiment} \\
\midrule
MSI   &       &    2.55** &    4.96* &    5.32* &       &    6.29*** &   11.49*** &   13.81*** &       &    3.84** &    6.87*** &    8.97*** &       &    2.65** &    5.03* &    5.99* \\
[0.5em]
SII$_{\text{is}}$ &       &   -2.81 &   -7.13 &  -16.42 &       &   -0.73 &   -2.98 &  -13.11 &       &    2.02** &    3.87*** &    2.84*** &       &   -3.75 &   -9.69 &  -22.76 \\
[0.5em]
SII$_{\text{oos}}$ &       &    0.18** &    3.64*** &    4.83*** &       &    0.97*** &    6.10*** &    6.18*** &       &    0.79** &    2.55*** &    1.41** &       &   -1.99 &   -0.01 &   -1.82 \\
[0.5em]
PLS   &       &    0.50 &    1.71** &    1.83* &       &    3.16*** &    7.50*** &    8.52*** &       &    2.67** &    5.89*** &    6.82*** &       &    0.26 &    1.47 &    0.72 \\
[0.5em]
PLS$_{\text{orth}}$ &       &    0.94* &    1.11* &   -2.24 &       &    3.38*** &    6.75*** &    4.55*** &       &    2.46** &    5.71*** &    6.65*** &       &    0.49 &    0.68 &   -3.58 \\
[0.5em]
AI &       &    7.35*** &   19.29*** &   19.61** &       &   11.16*** &   25.30*** &   28.30*** &       &    4.11** &    7.44*** &   10.81*** &       &    5.69*** &   17.69*** &   17.75** \\
[0.5em]
IVS   &       &    5.09*** &   11.31** &    8.24** &       &   12.29*** &   22.96*** &   22.26*** &       &    7.58*** &   13.13*** &   15.28*** &       &    4.93*** &   14.97** &    7.84** \\
[0.5em]
sntm  &       &   -1.59 &   -5.19 &  -11.23 &       &    2.27*** &    3.98*** &    6.25*** &       &    3.80*** &    8.72*** &   15.71*** &       &   -0.99 &   -3.16 &   -8.58 \\
\bottomrule
\end{tabular}%
}
\label{table: out-of-sample bivariate skewness cross-section sentiment variance-related}%
\end{table}%

\begin{table}[t!]
\centering
\caption*{\footnotesize Table \ref{table: out-of-sample bivariate skewness cross-section sentiment variance-related} \normalfont -- \textit{Continued}}
\footnotesize
\renewcommand{\arraystretch}{1}
\resizebox{1\textwidth}{!}{  
\begin{tabular}{lccccccccccccccccc}
\toprule
&       & \multicolumn{3}{c}{Historical average (univariate)}       &       & \multicolumn{3}{c}{Historical average (bivariate)}       &       & \multicolumn{3}{c}{Alternative predictor}       &       & \multicolumn{3}{c}{$\RSSevenFive$}  \\
\cmidrule{3-5}\cmidrule{7-9}\cmidrule{11-13}\cmidrule{15-17}          &       & $h = 1$     & $h = 3$     & $h = 6$     &       & $h = 1$     & $h = 3$     & $h = 6$     &       & $h = 1$     & $h = 3$     & $h = 6$     &       & $h = 1$     & $h = 3$     & $h = 6$ \\
\midrule
$\RSSevenFive$ &       &    2.91*** &    6.12*** &    7.85*** &       &       &       &       &       &       &       &       &       &       &       &  \\
\midrule
\multicolumn{17}{l}{Panel D: Variance} \\
\midrule
Vm    &       & -5.09 & 4.02  & -18.38 &       &    4.51*** &   17.01*** &   -1.33 &       &    9.13*** &   13.54*** &   14.41*** &       &   -1.52 &    8.09 &  -18.41 \\
[0.5em]
Vvw   &       & 1.67  & 6.17  & -11.93 &       &    8.25*** &   17.69*** &    6.68*** &       &    6.69*** &   12.28*** &   16.62*** &       &    2.45 &    8.84 &   -9.06 \\
[0.5em]
Vew   &       & 1.09  & 1.74  & -13.29 &       &    7.15*** &   12.24*** &    5.74*** &       &    6.12*** &   10.68*** &   16.80*** &       &    1.28 &    2.81 &  -10.15 \\
[0.5em]
FVF   &       & -1.24 & -2.04 & -7.87 &       &    8.22*** &   10.65*** &    9.94*** &       &    9.35*** &   12.44*** &   16.51*** &       &    1.17** &    0.35** &   -7.54 \\
[0.5em]
impvar &       & -2.53 & -5.31 & -11.53 &       &    1.16*** &    3.11*** &   -3.74 &       &    3.60*** &    7.99*** &    6.98*** &       &   -1.80 &   -3.20 &  -12.58 \\
[0.5em]
rsvix &       & -3.80  & -8.88 & -5.00    &       &   -0.12 &   -2.00 &    2.34*** &       &    3.54*** &    6.32*** &    6.99*** &       &   -3.12 &   -8.65 &   -5.99 \\
[0.5em]
vp &       & -3.94 &    1.36* & 0.27  &       &   -0.26 &   10.12*** &   13.20*** &       &    3.54*** &    8.88*** &   12.97*** &       &   -4.21 &    0.34* &   -0.08 \\
[0.5em]
vp$_{\text{bad}}$ &       & -4.37 & -6.92 & -0.49 &       &   -0.09 &    0.91*** &   11.24*** &       &    4.09*** &    7.32*** &   11.67*** &       &   -4.04 &   -9.87 &   -2.35 \\
[0.5em]
cvp &       &    2.45* &    1.65** &    2.36* &       &    8.50*** &   14.90*** &   19.94*** &       &    6.20*** &   13.47*** &   18.00*** &       &    4.89** &    5.63*** &    7.68*** \\
[0.5em]
cvp$_{\text{bad}}$ &       & 0.16  & -2.64 & -0.86 &       &    4.68*** &    6.98*** &   16.76*** &       &    4.52*** &    9.37*** &   17.47*** &       &    0.92 &   -3.15 &    4.01** \\
\midrule
\multicolumn{17}{l}{Panel E: Other macroeconomic and financial variables} \\
\midrule
dp &       &   -4.01 &  -20.30 &  -29.90 &       &   -1.42 &  -15.62 &  -37.07 &       &    2.49** &    3.89*** &   -5.53 &       &   -4.46 &  -23.15 &  -48.76 \\
[0.5em]
dy    &       &   -3.07 &  -14.97 &  -22.91 &       &   -0.01 &   -8.53 &  -22.28 &       &    2.97** &    5.61*** &    0.52*** &       &   -3.01 &  -15.60 &  -32.70 \\
[0.5em]
ep    &       &   -4.28 &  -20.32 &  -38.21 &       &   -0.97 &  -10.25 &  -20.39 &       &    3.17*** &    8.36*** &   12.90*** &       &   -4.00 &  -17.44 &  -30.65 \\
[0.5em]
de    &       &   -4.50 &  -32.94 &  -69.78 &       &   -1.22 &  -22.23 &  -47.70 &       &    3.14*** &    8.05*** &   13.00*** &       &   -4.25 &  -30.20 &  -60.29 \\
[0.5em]
svar  &       &  -11.54 &   -5.95 &  -15.27 &       &   -8.20 &    0.18*** &   -8.58 &       &    3.00*** &    5.78*** &    5.81*** &       &  -11.44 &   -6.32 &  -17.83 \\
[0.5em]
bm    &       &    1.62** &    5.63** &   12.83*** &       &    3.89*** &   11.09*** &   18.43*** &       &    2.31** &    5.79*** &    6.43*** &       &    1.01* &    5.30** &   11.48*** \\
[0.5em]
ntis  &       &   -1.17 &   -4.87 &  -11.69 &       &    1.46** &    1.19*** &   -0.46 &       &    2.60*** &    5.78*** &   10.05*** &       &   -1.49 &   -5.25 &   -9.02 \\
[0.5em]
tbl   &       &   -0.02 &   -0.92 &   -1.28 &       &    2.64*** &    5.20*** &    8.30*** &       &    2.66*** &    6.06*** &    9.45*** &       &   -0.28 &   -0.97 &    0.48** \\
[0.5em]
lty   &       &   -2.12 &   -6.39 &   -9.25 &       &   -0.08 &   -1.35 &   -5.00 &       &    2.00** &    4.73*** &    3.89*** &       &   -3.08 &   -7.96 &  -13.95 \\
[0.5em]
ltr   &       &   -0.98 &   -1.74 &   -0.58 &       &    1.69** &    4.88*** &    7.36*** &       &    2.64*** &    6.51*** &    7.90*** &       &   -1.26 &   -1.31 &   -0.53 \\
[0.5em]
tms   &       &   -0.64 &   -2.07 &   -3.87 &       &    2.20** &    4.09*** &    5.77*** &       &    2.83*** &    6.03*** &    9.28*** &       &   -0.73 &   -2.16 &   -2.26 \\
[0.5em]
dfy   &       &   -3.37 &  -25.99 &  -78.56 &       &   -0.05 &  -17.89 &  -61.22 &       &    3.21*** &    6.43*** &    9.71*** &       &   -3.05 &  -25.57 &  -74.97 \\
[0.5em]
dfr   &       &   -6.23 &   -7.15 &   -4.76 &       &   -1.39 &    3.17*** &    5.93*** &       &    4.56*** &    9.63*** &   10.20*** &       &   -4.42 &   -3.14 &   -2.09 \\
[0.5em]
infl  &       &   -1.15 &   -3.94 &    0.82* &       &    1.55** &    1.65*** &    8.59*** &       &    2.67*** &    5.38*** &    7.84*** &       &   -1.40 &   -4.76 &    0.80* \\
\midrule
ogap  &       &   -2.17 &  -10.90 &    1.50** &       &   -0.54 &   -7.70 &    9.18*** &       &    1.59** &    2.88** &    7.79*** &       &   -3.55 &  -14.71 &    1.43** \\
[0.5em]
ndrbl &       &   -4.08 &  -11.54 &  -13.15 &       &   -0.88 &   -2.63 &    1.79*** &       &    3.08*** &    7.99*** &   13.20*** &       &   -3.90 &   -9.32 &   -6.59 \\
[0.5em]
tchi  &       &   -0.69 &   -1.42 &   -1.22 &       &    2.49*** &    4.94*** &    7.37*** &       &    3.15*** &    6.27*** &    8.49*** &       &   -0.43 &   -1.25 &   -0.53 \\
[0.5em]
dtoy  &       &   -2.44 &  -11.83 &  -16.16 &       &    0.84** &   -4.37 &   -6.42 &       &    3.20*** &    6.67*** &    8.38*** &       &   -2.13 &  -11.17 &  -15.49 \\
[0.5em]
dtoat &       &   -2.89 &  -10.38 &  -10.65 &       &    0.99** &   -0.62 &    2.98** &       &    3.78*** &    8.85*** &   12.32*** &       &   -1.97 &   -7.17 &   -5.29 \\
[0.5em]
wtexas &       &   -3.62 &   -2.23 &   -1.61 &       &   -0.28 &    4.87*** &    6.88*** &       &    3.23*** &    6.94*** &    8.36*** &       &   -3.28 &   -1.33 &   -1.06 \\
[0.5em]
lzrt  &       &   -1.65 &   -5.43 &   -7.07 &       &    1.29*** &    0.21*** &   -0.29 &       &    2.89*** &    5.35*** &    6.33*** &       &   -1.67 &   -6.29 &   -8.84 \\
[0.5em]
ygap  &       &   -4.28 &  -20.34 &  -38.16 &       &   -1.01 &  -10.36 &  -20.55 &       &    3.14*** &    8.29*** &   12.75*** &       &   -4.03 &  -17.55 &  -30.82 \\
\bottomrule
\end{tabular}%
}
\label{table: out-of-sample bivariate goyal welth}%
\end{table}%
   

Table \ref{table: out-of-sample bivariate skewness cross-section sentiment variance-related} presents the out-of-sample statistics. The second through fourth columns correspond to univariate regressions as the main model and the historical average forecast as the benchmark. The fifth through the thirteenth columns present the cases with bivariate regressions as the main model and the three forecasts as the benchmark. Several insights can be drawn from these results. First, the strong in-sample predictive power of several alternative predictors translates into significant out-of-sample performance (relative to the historical average as the benchmark). Most prominently, the forward skewness factor, attention index, and implied volatility spread generate the highest and strongly significant $R_{OOS}^2$ statistics. Our findings are not inconsistent with \cite{welch2008comprehensive} in the sense that, except for the book-to-market ratio, none of the macroeconomic and financial variables generate positive and significant $R_{OOS}^2.$

Second, incorporating skewness dispersion into the regression model significantly enhances out-of-sample forecasting performance relative to the historical mean. Notably, all skewness-related predictors, four of the six stock cross-section variables, and nearly all sentiment and variance-related measures yield positive and statistically significant $R_{OOS}^2$ values. Interestingly, half of the other macroeconomic and financial variables, coupled with skewness dispersion, produce positive out-of-sample statistics at least for two predictive horizons. 

Third, the eighth through tenth columns compare the out-of-sample performance of the bivariate models with that of the univariate regression using the alternative predictor. Except for the dividend-to-price ratio and a 6-month horizon, skewness dispersion provides a significant incremental contribution. Conversely, only a small number of alternative predictors improve the model's accuracy when skewness dispersion alone is used. These include FSF and fbm from the skewness and stock cross-section categories; three sentiment measures (MSI, AI, and IVS); cvp from the variance group; and only the book-to-market ratio from the last set.

\subsubsection{Forecast encompassing tests}

We now implement the forecast encompassing test, which directly compares the information contained in the prediction based on skewness dispersion with that based on control variables. Specifically, we form a convex combination of predicted values from univariate regressions using $\RSSevenFive$ $\left(\widehat{r}_{t, t+ h}^{\RSSevenFive}\right)$ and a selected control $c$ $\left(\hat{r}_{t, t+ h}^c\right):$
\begin{equation*}
\hat{r}_{t, t+h}^*= \left( 1 - \lambda \right) \hat{r}_{t, t + h}^c +  \lambda  \hat{r}_{t, t+ h}^{\RSSevenFive} \quad \text{for} \quad 0 \le \lambda \le 1.
\end{equation*}
The value of $\lambda$ measures the contribution of $\hat{r}_{t, t+ h}^{\RSSevenFive}$ to the optimal combination forecast. If $\lambda = 0$, then this contribution is excluded from $\hat{r}_{t, t+h}^*$; that is, skewness dispersion does not contain any incremental information for forecasting average market returns beyond the information in the chosen predictor. In this case, the control variable encompasses the predictive regression forecast based on $\RSSevenFive.$ Alternatively, if $\lambda = 1$, skewness dispersion encompasses the information in the alternative predictor. For $\lambda \in (0,1),$ the optimal combination includes both forecasts, and the exact value of $\lambda$ measures the strength of the skewness dispersion forecast. 

\begin{table}[t!]
\centering
\caption{Out-of-sample forecast encompassing test}
\begin{minipage}{1.0\textwidth} 
\footnotesize This table reports the out-of-sample results for the forecast encompassing test. The out-of-sample estimation is based on an expanding window from December 2005 to December 2022. The dependent variable is the average monthly market excess returns in logarithm over the $h$-month horizon. The independent variable is the monthly skewness dispersion measure $\RSSevenFive$ (calculated as the mean of daily estimates in the last week of the month) and control variables, one at a time. The forecast encompassing test is $H_0: \lambda = 0, H_1: \lambda > 0,$ in which $\lambda$ is the weight of a univariate predictive forecast based on $\RSSevenFive$ in a convex combination of a univariate predictive forecast based on $\RSSevenFive$ and a univariate predictive forecast based on alternative predictors (not $\RSSevenFive$), one at a time. 
Following \cite{harvey1998tests}, we estimate the optimal weight $\lambda$ from the regression of non-$\RSSevenFive$-based forecast errors on the difference between non-$\RSSevenFive$-based and $\RSSevenFive$-based forecast errors. We report the t-statistic based on the t-distribution of the one-tail hypothesis test in brackets, following Equation (4) in \cite{harvey1998tests}. $^{*},^{**},$ and $^{***}$ indicate significance at the 10\%, 5\%, and 1\% levels, respectively.
\medskip
\end{minipage}
\footnotesize
\begin{tabular}{lccccclcccc}
\toprule
&       & $h = 1$     & $h = 3$     & $h = 6$     &       &       &       & $h = 1$     & $h = 3$     & $h = 6$ \\
\midrule
\multicolumn{11}{l}{Panel A: Skewness} \\
\midrule
Skm   &       &    1.00*** &    0.81*** &    0.91*** &       & Skew   &       &    0.92*** &    0.73*** &    0.88*** \\
 &       & [   3.36] & [   3.81] & [   3.70] &       &  &       & [   2.95] & [   3.60] & [   3.62] \\
[1.0em]
Skvw &       &    1.00*** &    0.66*** &    0.84*** &       & FSF &       &    0.55** &    0.31** &    0.62*** \\
 &       & [   2.78] & [   3.49] & [   3.64] &       &  &       & [   2.07] & [   1.68] & [   2.60] \\
\midrule
\multicolumn{11}{l}{Panel B: Stock cross-section} \\
\midrule
DISP &       &    0.65*** &    0.62*** &    0.76*** &       & avgcor &       &    0.87*** &    0.78*** &    0.72*** \\
 &       & [   2.45] & [   3.32] & [   3.05] &       &  &       & [   3.13] & [   4.00] & [   3.92] \\
[1.0em]
disag &       &    0.85*** &    0.82*** &    0.87*** &       & tail &       &    0.75*** &    0.72*** &    0.70*** \\
 &       & [   3.10] & [   3.42] & [   2.51] &       &  &       & [   2.87] & [   3.88] & [   3.73] \\
[1.0em]
rdsp &       &    0.79*** &    0.83*** &    0.85*** &       & fbm &       &    0.61** &    0.59*** &    0.64*** \\
 &       & [   2.48] & [   2.87] & [   2.91] &       &  &       & [   2.25] & [   3.14] & [   3.43] \\
\midrule
\multicolumn{11}{l}{Panel C: Sentiment} \\
\midrule
MSI &       &    0.60* &    0.58** &    0.61** &       & PLS$_{\text{orth}}$ &       &    0.63** &    0.64*** &    0.72*** \\
 &       & [   1.51] & [   1.89] & [   2.09] &       &  &       & [   2.29] & [   3.41] & [   3.42] \\
[1.0em]
SII$_{\text{is}}$ &       &    0.84*** &    0.77*** &    0.78*** &       & AI &       &    0.44* &    0.31* &    0.39** \\
 &       & [   3.67] & [   4.80] & [   4.92] &       &  &       & [   1.53] & [   1.58] & [   2.14] \\
[1.0em]
SII$_{\text{oos}}$ &       &    0.64*** &    0.53*** &    0.52*** &       & IVS &       &    0.56* &    0.49** &    0.63*** \\
 &       & [   2.59] & [   3.08] & [   2.64] &       &  &       & [   1.47] & [   2.28] & [   2.38] \\
[1.0em]
PLS   &       &    0.67*** &    0.62*** &    0.63*** &       & sntm &       &    0.83*** &    0.85*** &    1.00*** \\
 &       & [   2.51] & [   3.53] & [   3.47] &       &  &       & [   2.96] & [   4.20] & [   4.70] \\
\bottomrule
\end{tabular}%
\label{table: out-of-sample lambda skewness cross-section sentiment variance-related}%
\end{table}%

\begin{table}[t!]
\centering
\caption*{\footnotesize Table \ref{table: out-of-sample lambda skewness cross-section sentiment variance-related} \normalfont -- \textit{Continued}}
\footnotesize
\begin{tabular}{lccccclcccc}
\toprule
&       & $h = 1$     & $h = 3$     & $h = 6$     &       &       &       & $h = 1$     & $h = 3$     & $h = 6$ \\
\midrule
\multicolumn{11}{l}{Panel D: Variance} \\
\midrule
Vm &       &    0.73** &    0.60*** &    0.92** &       & rsvix &       &    0.83*** &    0.88*** &    0.77*** \\
 &       & [   2.01] & [   2.89] & [   2.24] &       &  &       & [   2.75] & [   3.16] & [   2.98] \\
[1.0em]
Vvw &       &    0.68** &    0.56*** &    0.89** &       & vp &       &    1.00*** &    0.71*** &    0.78*** \\
 &       & [   2.30] & [   2.79] & [   2.29] &       &  &       & [   3.35] & [   3.38] & [   3.80] \\
[1.0em]
Vew &       &    0.71** &    0.64*** &    0.97*** &       & vp$_{\text{bad}}$ &       &    0.96*** &    1.00*** &    0.81*** \\
 &       & [   2.14] & [   2.66] & [   2.53] &       &  &       & [   2.83] & [   3.69] & [   3.64] \\
[1.0em]
FVF &       &    0.68*** &    0.67*** &    0.83*** &       & cvp &       &    0.58** &    0.70*** &    0.70*** \\
 &       & [   2.93] & [   2.95] & [   3.12] &       &  &       & [   1.94] & [   4.20] & [   4.02] \\
[1.0em]
impvar &       &    0.77*** &    0.76*** &    0.82*** &       & cvp$_{\text{bad}}$ &       &    0.75** &    0.85*** &    0.80*** \\
 &       & [   2.64] & [   2.87] & [   2.76] &       &  &       & [   2.19] & [   3.33] & [   3.99] \\
\midrule
\multicolumn{11}{l}{Panel E: Other macroeconomic and financial variables} \\
\midrule
dp   &       &    0.90*** &    1.00*** &    1.00*** &       & tbl &       &    0.69*** &    0.67*** &    0.66*** \\
  &       & [   2.81] & [   3.10] & [   2.51] &       &   &       & [   2.72] & [   3.85] & [   3.38] \\
[1.0em]
dy &       &    0.82*** &    0.95*** &    0.97*** &       & lty &       &    0.72*** &    0.70*** &    0.68*** \\
  &       & [   2.94] & [   3.31] & [   2.66] &       &   &       & [   3.41] & [   4.27] & [   4.01] \\
[1.0em]
ep &       &    0.87** &    1.00** &    1.00*** &       & ltr &       &    0.79*** &    0.72*** &    0.69*** \\
  &       & [   2.11] & [   2.30] & [   2.38] &       &   &       & [   2.97] & [   3.84] & [   3.60] \\
[1.0em]
de &       &    0.77** &    0.97** &    1.00** &       & tms &       &    0.76*** &    0.73*** &    0.75*** \\
  &       & [   1.80] & [   1.77] & [   1.76] &       &   &       & [   2.77] & [   3.96] & [   3.95] \\
[1.0em]
svar &       &    0.84** &    0.71*** &    0.81*** &       & dfy &       &    0.72** &    0.84** &    1.00** \\
  &       & [   2.10] & [   2.58] & [   2.43] &       &   &       & [   2.16] & [   2.01] & [   1.70] \\
[1.0em]
bm &       &    0.57** &    0.51*** &    0.44*** &       & dfr &       &    1.00*** &    0.81*** &    0.74*** \\
  &       & [   2.29] & [   3.17] & [   2.79] &       &   &       & [   3.03] & [   3.72] & [   3.54] \\
[1.0em]
ntis &       &    0.75*** &    0.75*** &    0.79*** &       & infl &       &    0.79*** &    0.79*** &    0.65*** \\
  &       & [   2.65] & [   3.29] & [   3.03] &       &   &       & [   2.85] & [   4.22] & [   3.57] \\
\midrule
  &       &       &       &       &       &  &       &       &       &  \\
ogap &       &    0.71** &    0.79*** &    0.59*** &       & dtoat &       &    0.81*** &    0.82*** &    0.76*** \\
  &       & [   1.97] & [   2.66] & [   3.45] &       &   &       & [   3.47] & [   4.68] & [   3.98] \\
[1.0em]
ndrbl &       &    0.88*** &    0.91*** &    0.88*** &       & wtexas &       &    0.92*** &    0.70*** &    0.71*** \\
  &       & [   2.66] & [   3.24] & [   4.12] &       &   &       & [   2.98] & [   3.88] & [   3.60] \\
[1.0em]
tchi &       &    0.69*** &    0.66*** &    0.65*** &       & lzrt &       &    0.69*** &    0.69*** &    0.72*** \\
  &       & [   2.44] & [   2.97] & [   2.79] &       &   &       & [   3.18] & [   4.06] & [   3.80] \\
[1.0em]
dtoy &       &    0.83*** &    0.92*** &    0.96*** &       & ygap &       &    0.87** &    1.00** &    1.00*** \\
  &       & [   2.53] & [   2.91] & [   2.94] &       &   &       & [   2.11] & [   2.32] & [   2.40] \\
\bottomrule
\end{tabular}%
\label{table: out-of-sample lambda goyal welth}%
\end{table}%

Table \ref{table: out-of-sample lambda skewness cross-section sentiment variance-related} reports the estimate of $\lambda$ for each control variable considered and tests its significance following the procedure in \cite*{harvey1998tests}. We can reject the null hypothesis $\lambda = 0$ at the 10\% (5\%) confidence level in all (a vast majority of) control-horizon combinations. Thus, skewness dispersion contributes significantly to the optimal forecast. Except for the attention index (AI), this contribution exceeds 0.5; that is, the optimal forecast is mainly determined by the information contained in the skewness dispersion measure. Compared to the out-of-sample results reported in Table \ref{table: out-of-sample bivariate skewness cross-section sentiment variance-related}, the forecast encompassing test provides more substantial evidence for the superior information content of skewness dispersion relative to existing predictors. 

\subsection{Asset allocation}

This section evaluates the predictive power of skewness dispersion and other predictors from the perspective of an investor who allocates his wealth between the stock market and risk-free bill \citep{ferreira2011forecasting, rapach2016short}. Given an investment horizon of $h$ months, the investor allocates the proportions $w_t^h$ and $1 - w_t^h$ of his wealth to the aggregate equity market and the risk-free asset at the end of period $t$. The weight $w_t^h$ is defined as:
\begin{equation}
w_t^h = \frac{1}{\gamma} \frac{\hat{r}_{t, t + h}}{\hat{\sigma}_{t, t+h}^2},
\label{eq:weights}
\end{equation}
where $\gamma$ is the investor's coefficient of relative risk aversion, $\hat{r}_{t, t + h}$ is the forecast of return based on the predictive model, and $\hat{\sigma}_{t, t+h}^2$ is the forecast of the return variance. We assume that the weights are rebalanced at the same frequency as the forecast horizon. We estimate the univariate regression to generate the return forecast and compute realized volatility as a proxy for the volatility forecast. Consistent with the out-of-sample tests, the out-of-sample portfolio formation is based on the expanding window from December 2005 to December 2022, in which the forecast horizon and rebalancing frequency equal $h.$ We set the coefficient of relative risk aversion to three and restrict $w_t$ to the interval from $-0.5$ and $1.5$ to obtain well-behaved portfolio weights, common modeling choices in the literature \citep{campbell2008predicting,rapach2016short,han2021information}

Asset allocation utilizing Equation (\ref{eq:weights}) yields the following certainty equivalent:
\begin{equation}
CER = \overline{R}_p - 0.5 \gamma \sigma_p^2,
\end{equation}
where $\overline{R}_p$ and $\sigma_p^2$ are the mean and variance of the portfolio return. We can interpret the CER as the risk-free return an investor would be willing to accept in place of holding the risky portfolio. We then calculate the CER for an investor using the historical average as a benchmark forecast and define the CER gain as the annualised difference between the CER for an investor using the predictive regression forecasts to determine asset allocation and the CER for an investor using the historical average. The annualised CER gain can be interpreted as the amount an investor would be willing to pay to obtain the predictive regression forecast instead of investing based on the historical average. In addition to the CER gain, we report the Sharpe ratios of the corresponding portfolios.

\begin{sidewaystable}[htbp]
\centering
\caption{Out-of-sample certainty equivalent return gains and Sharpe ratios}
\begin{minipage}{1\textwidth} 
\footnotesize This table reports the annualized certainty equivalent return (CER) gain and the annualized Sharpe ratio (SR) for a mean-variance investor with a relative risk aversion of three who allocates between the value-weighted market excess returns and the risk-free rate using a univariate predictive forecast based on a particular variable. The optimal equity weights are constrained to the interval [-0.5, 1.5]. The prevailing mean corresponds to the investor using the historical average as a benchmark forecast. Buy and hold corresponds to the passive investment in the market portfolio. Consistent with the out-of-sample tests, the out-of-sample portfolio formation is based on the expanding window from December 2005 to December 2022, in which the forecast horizon and rebalancing frequency equal $h.$
\medskip
\end{minipage}
\footnotesize
\renewcommand{\arraystretch}{1}
\resizebox{1\textwidth}{!}{  
\begin{tabular}{lrcccccccrlrccccccc}
\toprule
Predictor &       & \multicolumn{3}{c}{CER} &       & \multicolumn{3}{c}{SR} &       &       &       & \multicolumn{3}{c}{CER} &       & \multicolumn{3}{c}{SR} \\
\cmidrule{3-5}\cmidrule{7-9}\cmidrule{13-15}\cmidrule{17-19}          &       & $h=1$   & $h=3$   & $h=6$   &       & $h=1$   & $h=3$   & $h=6$   &       &       &       & $h=1$   & $h=3$   & $h=6$   &       & $h=1$   & $h=3$   & $h=6$ \\
\midrule
Buy and hold &       & 2.40  & 2.58  & 3.29  &       & 0.57  & 0.60  & 0.64  &       & \multicolumn{4}{l}{Prevailing mean}       &       &       & 0.42  & 0.52  & 0.47 \\
[1em]
$\RSSevenFive$ &       & 8.36  & 3.95  & 4.45  &       & 0.89  & 0.69  & 0.78  &       &       $\RSNineZero$ &       & 8.16  & 4.69  & 6.30  &       & 0.87  & 0.74  & 0.89\\
[0.5em]
$\RSEightZero$ &       & 8.56  & 3.94  & 4.99  &       & 0.90  & 0.69  & 0.81  &       &  $\RSNineFive$ &       & 7.09  & 4.39  & 7.29  &       & 0.82  & 0.72  & 0.94 \\
[0.5em]
$\RSEightFive$ &       & 8.75  & 4.43  & 5.57  &       & 0.91  & 0.72  & 0.85  &       & &       &       &       &       &       &       &       &  \\
\midrule
\multicolumn{19}{l}{Panel A: Skewness} \\
\midrule
Skm   &       & -3.75 & -0.86 & -2.73 &       & -0.03 & 0.39  & 0.15  &       & Skew  &       & -0.22 & 3.52  & 0.07  &       & 0.40  & 0.74  & 0.55 \\
[0.5em]
Skvw  &       & -4.69 & 4.08  & -2.20 &       & 0.03  & 0.90  & 0.21  &       & FSF   &       & -5.39 & 4.19  & -4.43 &       & 0.03  & 0.72  & -0.05 \\
\midrule
\multicolumn{19}{l}{Panel B: Stock cross-section} \\
\midrule
DISP  &       & 2.88  & 2.96  & 1.37  &       & 0.64  & 0.69  & 0.66  &       & avgcor &       & -4.97 & -3.10 & -2.85 &       & 0.10  & 0.20  & 0.19 \\
[0.5em]
disag &       & 1.47  & -1.18 & 0.99  &       & 0.53  & 0.35  & 0.54  &       & tail  &       & -0.07 & 0.64  & -0.59 &       & 0.41  & 0.53  & 0.38 \\
[0.5em]
rdsp  &       & 1.25  & -0.54 & -2.38 &       & 0.51  & 0.41  & 0.22  &       & fbm   &       & 3.70  & 2.27  & 0.11  &       & 0.68  & 0.68  & 0.46 \\
\midrule
\multicolumn{19}{l}{Panel C: Sentiment} \\
\midrule
MSI   &       & 4.00  & 3.90  & 1.07  &       & 0.72  & 0.77  & 0.53  &       & PLS$_{\text{orth}}$ &       & 3.61  & 1.43  & -3.14 &       & 0.71  & 0.60  & 0.22 \\
[0.5em]
SII$_{\text{is}}$ &       & 2.59  & 4.59  & 3.48  &       & 0.57  & 0.71  & 0.62  &       & AI$_{\text{PLS}}$ &       & 9.47  & 7.32  & 3.40  &       & 1.12  & 0.93  & 0.77 \\
[0.5em]
SII$_{\text{oos}}$ &       & 5.38  & 9.27  & 11.09 &       & 0.71  & 0.93  & 1.12  &       & IVS   &       & 2.67  & 4.33  & 0.42  &       & 0.58  & 0.71  & 0.54 \\
[0.5em]
PLS   &       & 1.92  & 1.35  & 0.43  &       & 0.56  & 0.58  & 0.46  &       & sntm  &       & -0.23 & -0.87 & -3.23 &       & 0.40  & 0.38  & 0.19 \\
\bottomrule
\end{tabular}%
}
\label{table: out-of-sample ceq sr 1}%
\end{sidewaystable}%
   
\begin{sidewaystable}[htbp]
\centering
\caption*{\footnotesize Table \ref{table: out-of-sample ceq sr 1} \normalfont -- \textit{Continued}}
\footnotesize
\renewcommand{\arraystretch}{1}
\resizebox{\textwidth}{!}{  
\begin{tabular}{lrrrrrrrrrlrrrrrrrr}
\toprule
Predictor &       & \multicolumn{3}{c}{CER} &       & \multicolumn{3}{c}{SR} &       &       &       & \multicolumn{3}{c}{CER} &       & \multicolumn{3}{c}{SR} \\
\cmidrule{3-5}\cmidrule{7-9}\cmidrule{13-15}\cmidrule{17-19}          &       & $h=1$   & $h=3$   & $h=6$   &       & $h=1$   & $h=3$   & $h=6$   &       &       &       & $h=1$   & $h=3$   & $h=6$   &       & $h=1$   & $h=3$   & $h=6$ \\
\midrule
\multicolumn{19}{l}{Panel D: Variance-related} \\
\midrule
Vm    &       & -2.28 & -0.60 & -2.91 &       & 0.21  & 0.44  & 0.12  &       & rsvix &       & -2.74 & -1.58 & -1.55 &       & 0.20  & 0.31  & 0.28 \\
[0.5em]
Vvw   &       & 2.40  & 1.09  & -2.50 &       & 0.72  & 0.56  & 0.19  &       & vp &       & -3.17 & 1.93  & 0.20  &       & 0.17  & 0.57  & 0.45 \\
[0.5em]
Vew   &       & 2.74  & -0.21 & -2.43 &       & 0.76  & 0.45  & 0.19  &       & vp$_{\text{bad}}$ &       & -2.39 & -1.72 & -1.02 &       & 0.21  & 0.29  & 0.33 \\
[0.5em]
FVF   &       & -1.82 & 0.89  & 0.61  &       & 0.25  & 0.53  & 0.47  &       & cvp   &       & 1.64  & -0.67 & 0.08  &       & 0.54  & 0.42  & 0.44 \\
[0.5em]
impvar &       & -0.51 & 0.49  & -1.95 &       & 0.37  & 0.54  & 0.24  &       & cvp$_{\text{bad}}$ &       & 1.81  & 0.34  & -0.43 &       & 0.58  & 0.52  & 0.39 \\
\midrule
\multicolumn{19}{l}{Panel E: Other macroeconomic and stock market variables} \\
\midrule
dp    &       & 0.90  & -4.40 & -2.48 &       & 0.47  & 0.19  & 0.29  &       & dfy   &       & 0.56  & 0.79  & 0.46  &       & 0.46  & 0.58  & 0.51 \\
[0.5em]
dy    &       & 1.85  & -4.05 & -1.41 &       & 0.54  & 0.22  & 0.35  &       & dfr   &       & -4.54 & -2.60 & -0.01 &       & 0.08  & 0.21  & 0.41 \\
[0.5em]
ep    &       & 1.01  & 0.66  & -0.19 &       & 0.50  & 0.53  & 0.43  &       & infl  &       & -1.87 & -1.85 & 1.24  &       & 0.26  & 0.28  & 0.59 \\
[0.5em]
de    &       & 1.64  & 1.24  & 0.83  &       & 0.56  & 0.59  & 0.51  &       & ogap  &       & 1.99  & 0.95  & 4.97  &       & 0.54  & 0.50  & 0.74 \\
[0.5em]
svar  &       & -1.15 & 1.25  & -0.28 &       & 0.33  & 0.62  & 0.42  &       & ndrbl &       & -1.15 & -0.51 & -2.80 &       & 0.32  & 0.42  & 0.19 \\
[0.5em]
bm    &       & 4.70  & 2.75  & 5.08  &       & 0.75  & 0.60  & 0.82  &       & tchi  &       & 2.10  & 1.85  & 3.79  &       & 0.55  & 0.58  & 0.75 \\
[0.5em]
ntis  &       & -1.64 & -1.53 & -0.80 &       & 0.28  & 0.31  & 0.36  &       & dtoy  &       & -2.15 & -2.72 & -2.17 &       & 0.24  & 0.20  & 0.22 \\
[0.5em]
tbl   &       & 3.39  & 2.32  & 1.54  &       & 0.65  & 0.62  & 0.51  &       & dtoat &       & -2.87 & -5.66 & -1.64 &       & 0.20  & -0.02 & 0.28 \\
[0.5em]
lty   &       & 2.42  & 1.16  & 2.11  &       & 0.57  & 0.52  & 0.55  &       & wtexas &       & -5.07 & -0.16 & 1.17  &       & 0.04  & 0.43  & 0.60 \\
[0.5em]
ltr   &       & -0.79 & -2.32 & 0.18  &       & 0.35  & 0.23  & 0.49  &       & lzrt  &       & 0.09  & 1.30  & 1.17  &       & 0.42  & 0.54  & 0.54 \\
[0.5em]
tms   &       & -1.18 & -0.48 & -1.68 &       & 0.32  & 0.44  & 0.27  &       & ygap  &       & 1.03  & 0.61  & -0.27 &       & 0.50  & 0.52  & 0.42 \\
\bottomrule
\end{tabular}%
}
\label{table: out-of-sample ceq sr 2}%
\end{sidewaystable}%

Table \ref{table: out-of-sample ceq sr 1} reports the CER gains and Sharpe ratios associated with univariate predictive regression forecasts based on each of the skewness dispersion measures and control variables. It also shows the performance results for the buy-and-hold strategy, which consists of a passive investment in the market portfolio, and the prevailing mean, which corresponds to an investor using the historical average as a benchmark. 

At the 1-month horizon, the CER gains from using the skewness-dispersion-based forecast range from 709 basis points for $\RSNineFive$ to hefty 875 basis points for $\RSEightFive$. Compared with other predictors, only the attention index shows a larger CER gain of 947 basis points. At the 3-month horizon, the CER gains tend to increase with the range of the cross-sectional skewness dispersion: from around 395 basis points for $\RSSevenFive$ up to 469 basis points for $\RSNineZero$. Only the short interest and attention indices produce a superior performance of 927 and 732 basis points, whereas several other predictors generate comparable CER gains. At the 6-month horizon, the CER gains associated with skewness dispersion increase from 445 basis points for $\RSSevenFive$ to 729 basis points for $\RSNineFive$. Only the out-of-sample short interest index shows a larger CER gain of 1109 basis points. The market portfolio underperforms the skewness dispersion strategies across all horizons.

\begin{figure}[t!]
\centering
\includegraphics[width=1.00\linewidth]{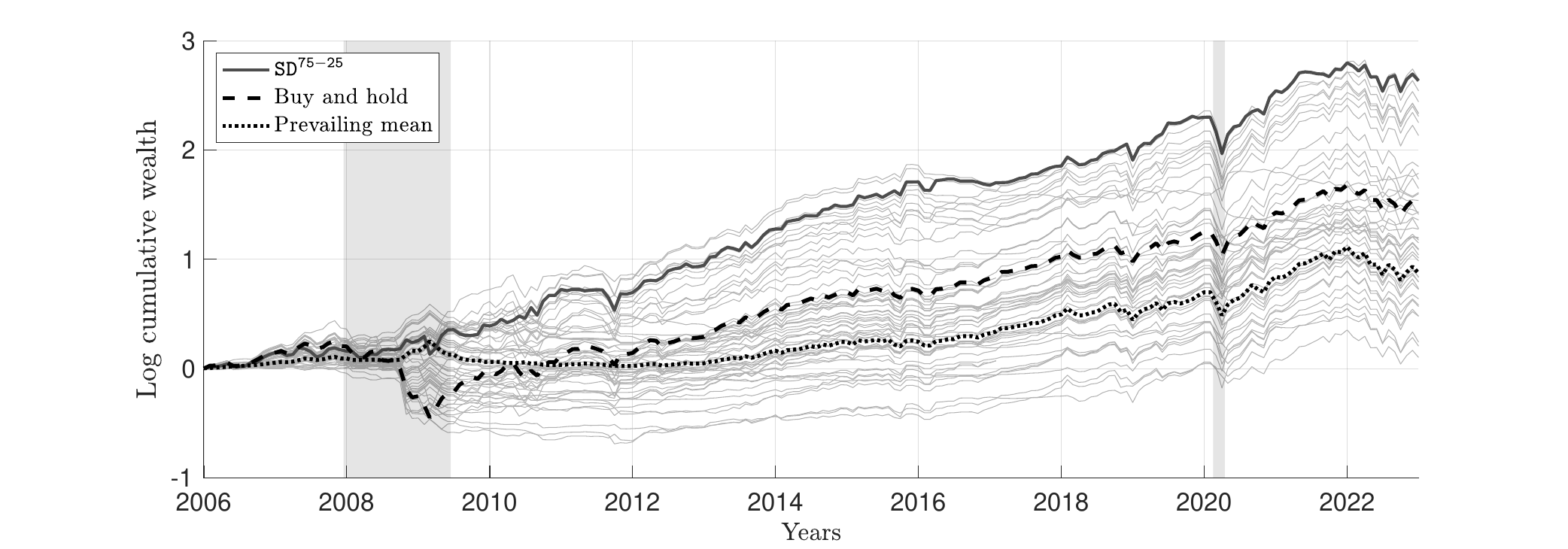}
\caption{Out-of-sample log cumulative wealth}
\begin{minipage}{\textwidth} \footnotesize This figure shows the logarithm of the cumulative wealth of a mean-variance investor with a relative risk aversion of three, assuming the initial investment is \$1 and all proceeds are reinvested. The investor allocates between value-weighted market excess returns and the risk-free rate using a univariate predictive forecast based on a given variable. The optimal equity weights are constrained to the range -0.5 to 1.5. For ease of reading, we highlight the investor's cumulative wealth using $\RSSevenFive$ (calculated as the mean of daily estimates in the last week of the month) or the historical average as the benchmark forecast and compare it with the buy-and-hold strategy. Buy and hold corresponds to a passive investment in the market portfolio. Other strategies are shown in light gray. Consistent with the out-of-sample statistical performance, the out-of-sample portfolio construction uses an expanding window from December 2005 to December 2022, with a one-month forecast horizon and rebalancing frequency. The shaded bars indicate NBER recession periods.
\end{minipage}
\label{figure: out of sample log cumulative wealth}
\end{figure}

Table \ref{table: out-of-sample ceq sr 1} demonstrates similar patterns in Sharpe ratio statistics across different predictive regressions. For instance, Sharpe ratios associated with skewness dispersion forecasts at the 1-month horizon range from 0.82 to 0.91. They are dominated only by a Sharpe ratio of 1.12 corresponding to the attention index, consistent with CER gains. The out-of-sample short interest index and the attention index have the highest Sharpe ratio of 0.93 at the 3-month horizon, followed by the Sharpe ratios for skewness dispersion measures from 0.69 to 0.74. Finally, the Sharpe ratios of the 6-month-ahead returns for portfolios constructed using skewness dispersion forecasts range from 0.78 to 0.94. Except for the out-of-sample short interest index, no other predictor yields a Sharpe ratio of at least 0.78 over the 6-month forecast horizon, again consistent with our findings based on CER gains. 

We also compute the log cumulative wealth generated by different strategies, allowing us to track portfolio gains over time. Figure \ref{figure: out of sample log cumulative wealth} compares the results across various forecast models. The $\RSSevenFive$ portfolio starts to outperform all benchmarks except for two portfolios in the aftermath of the Global Financial Crisis. It generates substantial cumulative wealth gains toward the end of the sample, outpacing virtually all other strategies. Thus, skewness dispersion forecasts tend to yield the highest CER gains and Sharpe ratios among a large set of alternative predictors. A few sentiment-based measures perform comparably, whereas other predictors strongly underperform. The skewness dispersion forecasts also achieve superior performance in terms of cumulative wealth.

\section{Interpretation of the predictive power of skewness dispersion}
\label{section: interpretation of the predictive power of skewness dispersion}

\subsection{Rational and behavioral channels}
\label{section: rational and behavioral channels}

We next investigate the economic mechanisms underlying the predictive power of skewness dispersion. In particular, we assess whether the documented return predictability is consistent with risk-based or behavioral explanations. Table \ref{table: correlations 1} reports contemporaneous correlations between ${\tt SD^{75-25}}$ and the variables listed in Table \ref{tab:variables}. If skewness dispersion proxies for risk, we expect to observe a negative correlation with ex-ante measures of market volatility or a positive correlation with skewness. Intuitively, if the negative relation between ${\tt SD^{75-25}}$ and future stock market returns reflects a time-varying risk premium, then high skewness dispersion states represent good times to investors (as they require lower expected returns), which should be accompanied by lower risk or higher skewness measures. 

Consistent with this interpretation, Table \ref{table: correlations 1} shows that the correlation with stock market option-implied variances (impvar and rsvix) and variance risk premia ($vp$ and $vp_{bad}$) is negative and significant at the 1\% level, whereas the correlation with realized variances of market returns (Vm and svar) is negative and significant at the 5\% level. Moreover, skewness dispersion exhibits an economically and statistically meaningful negative correlation with the average correlation of daily stock returns, which has been shown to relate to changes in aggregate risk. Specifically, \cite{pollet2010average} argue that, other things equal, an increase in aggregate risk is associated with stronger comovement in stock prices. Hence, higher aggregate risk can be revealed by a higher correlation between stocks. The observed negative relation, therefore, suggests that higher skewness dispersion is associated with lower aggregate risk. Meanwhile, the relationship between skewness dispersion and skewness measures is weaker and often insignificant. We find only limited evidence of a positive association with average stock skewness (Skew), and no meaningful relation with tail risk measures. Thus, we find that skewness dispersion is related to variation in aggregate risk, but not to crash risk.

We next examine several possible behavioral explanations. Under a sentiment-based view, high skewness dispersion should coincide with elevated investor sentiment and, consequently, lower subsequent returns. Consistent with this hypothesis, Table \ref{table: correlations 1} shows a positive and significant correlation with the two investor sentiment measures (PLS, PLS$_{\text{orth}}$) of \cite{huang2015investor} and the investor attention index (AI) of \cite{chen2022investor}.\footnote{It is noteworthy that the dataset of \cite{goyal2024comprehensive} includes the investor sentiment measure (sntm) of \cite{huang2015investor}, which is materially different from the time series posted on Guofu Zhou's website. We find that ${\tt SD^{75-25}}$ has a negative and significant correlation with sntm, which is inconsistent with the sentiment hypothesis of the behavioral explanation.} The former are based on the six individual investor sentiment proxies used by \cite{baker2006investor}. When we consider a composite sentiment measure of \cite{baker2006investor}, ${\tt SD}^{75-25}$ exhibits a positive but insignificant correlation of 0.07. This complements a negligible relationship with the manager sentiment index (MSI) of \cite{jiang2019manager}, which is based on corporate filings and earnings conference call transcripts. 

Our evidence indicates a positive relationship with several sentiment measures, as predicted, but the effects are generally modest and not always significant. We find the strongest association with investor attention, prompting us to better understand this link to investor expectations. For this reason, we consider other sentiment proxies, including bullish, neutral, and bearish measures, based on survey data from the American Association of Individual Investors (AAII). The AAII sentiment survey summarizes opinions from individual investors by asking them where the market is heading over the next six months. We find a strong positive (negative) correlation of ${\tt SD^{75-25}}$ with bullish (bearish) sentiment of 0.30 ($-0.28$), which is significant at the 1\% level. This suggests that skewness dispersion is unlikely to capture investor sentiments reflected in the established measures, and instead reflects investor beliefs directly observed in the survey data.

\begin{table}[t!]
\centering
\caption{Predictive power of skewness dispersion in different regimes} 
\begin{minipage}{\textwidth} 
\footnotesize This table reports the in-sample results of univariate predictive regressions estimated for different regimes. The dependent variable is the average monthly market excess returns in logarithm over the $h$-month horizon. The independent variable is a monthly skewness dispersion measure, $\RSSevenFive$, calculated as the mean of daily estimates from the last week of the month. Panel A reports the slope coefficients, \cite{newey1986simple} adjusted statistics based on $h-1$ lags in brackets, and adjusted $R^2$ of OLS regressions estimated on the subsamples of the whole period from December 2000 to December 2022 based on the investor sentiment of \cite{baker2006investor} and the AAII bullish-bearish spread. Panel B reports regression outputs using data from the months before, during, and after FOMC meetings, as well as data excluding FOMC meetings. $^{*},^{**},$ and $^{***}$ indicate significance at the 10\%, 5\%, and 1\% levels, respectively.
\medskip
\end{minipage}
\footnotesize
\begin{tabular}{lccccc}
\toprule
\multicolumn{6}{l}{Panel A: Sentiment and market expectations} \\
\midrule
&       & 1     & 3     & 6     & 12 \\
\midrule
&       & \multicolumn{4}{c}{High sentiment regime} \\
\cmidrule{3-6}    
$\RSSevenFive$ &       &  -0.076** &  -0.068** &  -0.063* &  -0.053** \\
      &       & [  -2.00] & [  -2.06] & [  -1.94] & [  -2.36] \\
$R^2_{\text{adj}}$ &       &    2.92 &    6.12 &    7.75 &    7.99 \\
\cmidrule{3-6}    
&       & \multicolumn{4}{c}{Low sentiment regime} \\
\cmidrule{3-6}    
$\RSSevenFive$ &       &  -0.111** &  -0.091*** &  -0.064*** &  -0.027*** \\
      &       & [  -2.14] & [  -4.74] & [  -4.73] & [  -4.28] \\
$R^2_{\text{adj}}$ &       &    4.51 &   11.06 &   15.64 &    9.96 \\
\midrule
&       & \multicolumn{4}{c}{High bullish-bearing regime} \\
\cmidrule{3-6}    
$\RSSevenFive$ &       &  -0.164*** &  -0.106*** &  -0.069*** &  -0.039* \\
      &       & [  -4.32] & [  -3.75] & [  -3.00] & [  -1.80] \\
$R^2_{\text{adj}}$ &       &   13.45 &   13.85 &   11.71 &    7.18 \\
\cmidrule{3-6}    
&       & \multicolumn{4}{c}{Low bullish-bearish regime} \\
\cmidrule{3-6}    
$\RSSevenFive$ &       &  -0.072 &  -0.079*** &  -0.073*** &  -0.052*** \\
      &       & [  -1.56] & [  -2.89] & [  -2.63] & [  -3.77] \\
$R^2_{\text{adj}}$ &       &    1.81 &    7.03 &   10.38 &    8.93 \\
\midrule
\multicolumn{6}{l}{Panel B: FOMC meetings} \\
\midrule
 &       & FOMC  & pre-FOMC & post-FOMC & non-FOMC \\
\midrule
$\RSSevenFive$ &       &  -0.133*** &  -0.108*** &  -0.080* &  -0.018 \\
      &       & [  -3.28] & [  -2.97] & [  -1.95] & [  -0.31] \\
$R^2_{\text{adj}}$ &       &    7.08 &    5.11 &    2.83 &    0.15 \\
\bottomrule
\end{tabular}%
\label{table: in-sample subsamples}%
\end{table}%

According to the mispricing hypothesis of the behavioral explanation, greater disagreement, coupled with short selling constraints, gives rise to overvaluation \citep{miller1977risk}. Consequently, high skewness dispersion should be contemporaneously related to high disagreement measures, predicting lower future market returns. Panel B in Table \ref{table: correlations 1}, however, shows insignificant correlations with all measures of investor and analyst disagreement. Interestingly, we find a positive and highly statistically significant correlation between skewness dispersion and aggregate short interest. If skewness dispersion indeed captures the overpricing of individual stocks that rational investors exploit, we should observe a more pronounced predictive power during periods of optimism.

We test this conjecture by examining the predictive power of ${\tt SD}^{75-25}$ conditional on investor sentiment and expectations. In our analysis, we employ investor sentiment of \cite{baker2006investor} and the AAII bullish-bearish spread. For each case, we split the sample into high and low states based on whether the measure is above or below the median, respectively. We then estimate the predictive regressions using the dates belonging to one of the two regimes. Table \ref{table: in-sample subsamples} shows mixed results. On the one hand, skewness dispersion is a stronger predictor of market returns during periods of low sentiment. On the other hand, the predictive power of skewness dispersion is stronger when investors become more optimistic, as measured by the bullish-bearish spread. It is noteworthy that the effect does not completely disappear in one of the regimes, indicating that mispricing alone cannot fully account for the observed predictability. Our preferred interpretation is that stronger correlations between AAII sentiment measures and skewness dispersion, coupled with more pronounced differences in predictive power across AAII-based regimes, indicate that investor expectations underlying the AAII sentiment measures align closer with the behavioral content of skewness dispersion.

\subsection{Information diffusion around FOMC announcements}

The strong correlation with aggregate short interest indicates that cross-sectional variation in firm-level skewness reflects total short selling in the economy. Short selling positions are arguably associated with the expectations of relatively sophisticated investors about future stock returns. If skewness dispersion indeed captures such expectations, its predictive power should be amplified during periods of concentrated information arrival, when the general public quickly reflects that information in asset prices. 

\cite*{badidi2026macroeconomic} study various macroeconomic news and document that FOMC announcements matter most for investors. Motivated by this insight, we investigate how monetary policy announcements, which represent salient and well-identified macroeconomic information events, affect the predictive power of skewness dispersion. We first partition the sample into months containing FOMC meetings, months immediately preceding and following these meetings, and months without FOMC activity. We then estimate one-month-ahead predictive regressions within each subsample. 

The results in Table \ref{table: in-sample subsamples} show that the predictive power of $\RSSevenFive$ is sharply concentrated around FOMC-related periods. In months with FOMC meetings, $\RSSevenFive$ strongly predicts next-month excess returns, with a coefficient of $-0.133$ (t-statistic $-3.28$) and an adjusted $R^2$ of $7.08\%$. In contrast, the predictive relation is economically negligible and statistically insignificant in non-FOMC months, with a coefficient of $-0.018$ (t-statistic $-0.31$) and an adjusted $R^2$ close to zero. Importantly, predictability is already present in the month preceding FOMC announcements, with a coefficient of $-0.108$ (t-statistic $-2.97$), consistent with investors taking positions ahead of major policy releases. The effect remains statistically significant, albeit much weaker, in the post-FOMC period, suggesting that prices adjust quickly following news releases.

In sum, the predictive power of skewness dispersion is not pervasive but instead concentrated in periods of heightened information arrival. The FOMC evidence complements the earlier results by pointing to an information-based mechanism for return predictability. We conclude that the observed return predictability reflects both the repricing of aggregate risk, consistent with a risk-based explanation, and the adjustment of valuation discrepancies, in line with a behavioral explanation, in response to major macroeconomic announcements that are revealed and processed in the market.

\subsection{Kurtosis dispersion and market return predictability}

Realized skewness measures the asymmetry in the return distribution, indicating whether positive or negative returns are more prevalent. Our primary analysis demonstrates that the widening range of realized skewness estimates in the cross-section negatively predicts subsequent market returns. A natural question arises: is the result driven exclusively by the asymmetry in the return distribution or by the overall tail heaviness? 

\begin{table}[t!]
\centering
\caption{Univariate regression results: kurtosis dispersion} 
\begin{minipage}{\textwidth} 
\footnotesize This table reports the in-sample (Panels A through D) and out-of-sample (Panel E) results of univariate predictive regressions. The dependent variable is the average monthly market excess returns in logarithm over the $h$-month horizon. The independent variable is a monthly kurtosis dispersion measure ${\tt KD}^{75 - 25},$ calculated as the mean (median) of daily kurtosis in the last week of the month. Panel A (B) reports the slope coefficients, \cite{newey1986simple} adjusted t-statistics based on $h-1$ lags in brackets, and adjusted $R^2$ of OLS regressions estimated on the sample period from December 2000 to December 2022 (excluding NBER recessions). Panel C reports the regression outputs based on the full sample and non-overlapping observations, following the procedure of \cite{martin2017expected}. The 10\%, 5\%, and 1\% critical values for the t-statistic are 1.645, 1.960, and 2.576. Panel D reports the slope coefficients and IVX-Wald statistics from the \cite{kostakis2015robust} estimation based on the full sample. The 10\%, 5\%, and 1\% critical values for the Wald statistic are 2.706, 3.842, and 6.634. Panel E reports the out-of-sample $R^2$ (in percentages) computed as $R^2_{oos} = 1 - MSFE_L/MSFE_B,$ where $MSFE_B$ is the mean squared forecast error for a benchmark forecast equal to the historical average and $MSFE_L$ is the mean squared forecast error for a predictive regression using a particular predictor. We test the hypothesis $H_0: MSFE_B \le MSFE_L, H_1: MSFE_B \geq MSFE_L$ (or equivalently $H_0: R_{OOS}^2 \le 0, H_1: R_{OOS}^2 \geq 0$) by implementing the \cite{clark2007approximately} test and reporting the resulting \cite{newey1986simple} adjusted t-statistic based on $h-1$ lags in brackets. The 10\%, 5\%, and 1\% critical values for the t-statistic (a one-sided test) are 1.282, 1.645, and 2.334. $^{*},^{**},$ and $^{***}$ indicate significance at the 10\%, 5\%, and 1\% levels, respectively.
\medskip
\end{minipage}
\footnotesize
\begin{tabular}{lcccccccccc}
\toprule
&       & \multicolumn{4}{c}{Mean}      &       & \multicolumn{4}{c}{Median} \\
\cmidrule{3-6}\cmidrule{8-11}                   
&       & $h = 1$     & $h = 3$     & $h = 6$     & $h = 12$    &       & $h = 1$     & $h = 3$     & $h = 6$     & $h = 12$ \\
\midrule
\multicolumn{11}{l}{Panel A: Full sample} \\
\midrule
${\tt KD}^{75 - 25}$ &       &  -0.034*** &  -0.022*** &  -0.017** &  -0.012*** &       &  -0.034*** &  -0.021*** &  -0.016** &  -0.011*** \\
 &       & [  -3.34] & [  -3.01] & [  -2.28] & [  -3.22] &       & [  -3.40] & [  -3.00] & [  -2.25] & [  -3.16] \\
$R^2_{\text{adj}}$ &       &    5.93 &    7.35 &    7.73 &    6.41 &       &    5.79 &    6.91 &    6.81 &    5.32 \\
\midrule
\multicolumn{11}{l}{Panel B: Excluding NBER recessions} \\
\midrule
${\tt KD}^{75 - 25}$ &       &  -0.025*** &  -0.012** &  -0.009* &  -0.008*** &       &  -0.027*** &  -0.013** &  -0.008 &  -0.007*** \\
 &       & [  -3.00] & [  -2.31] & [  -1.76] & [  -2.80] &       & [  -3.27] & [  -2.32] & [  -1.60] & [  -2.90] \\
$R^2_{\text{adj}}$ &       &    4.05 &    3.32 &    3.27 &    4.62 &       &    4.62 &    3.45 &    2.82 &    3.89 \\
\midrule
\multicolumn{11}{l}{Panel C: Non-overlapping observations} \\
\midrule
${\tt KD}^{75 - 25}$ &       &  -0.034*** &  -0.021* &  -0.017** &  -0.011* &       &  -0.034*** &  -0.021* &  -0.016** &  -0.011** \\
 &       & [  -3.34] & [  -1.95] & [  -1.99] & [  -1.92] &       & [  -3.40] & [  -1.91] & [  -2.04] & [  -2.34] \\
$R^2_{\text{adj}}$ &       &    5.93 &    7.33 &    8.40 &   10.98 &       &    5.79 &    8.21 &    7.71 &    8.91 \\
\midrule
\multicolumn{11}{l}{Panel D: IVX estimation} \\
\midrule
${\tt KD}^{75 - 25}$ &       &  -0.035*** &  -0.015*** &  -0.008*** &  -0.005*** &       &  -0.035*** &  -0.015*** &  -0.008*** &  -0.005*** \\
IVX-Wald &       & [  17.83] & [  14.42] & [  11.66] & [   7.63] &       & [  17.20] & [  13.89] & [  10.93] & [   6.87] \\
\midrule
\multicolumn{11}{l}{Panel E: Out-of-sample estimation} \\
\midrule
${\tt KD}^{75 - 25}$ &       &    4.91*** &    4.72*** &    2.78*** &    2.37** &       &    4.99*** &    5.57*** &    3.26*** &    1.48* \\
 &       & [   3.09] & [   3.08] & [   2.68] & [   2.26] &       & [   3.26] & [   3.31] & [   2.65] & [   1.56] \\
\bottomrule
\end{tabular}%
\label{table: univariate kurtosis}%
\end{table}%

We answer this question by investigating the relationship between the cross-sectional dispersion in firm-level kurtosis and stock market returns. Following the definition of skewness dispersion, we compute the daily realized kurtosis for each stock and calculate the inter-quartile range of the daily estimates. Then, we convert the daily kurtosis dispersion measure to a monthly time series by taking the mean (median) in the last week of the month. For brevity, we replicate the in-sample and out-of-sample predictive regressions with kurtosis dispersion.

Table \ref{table: univariate kurtosis} reports the univariate regression results. It shows a strong, negative relationship between kurtosis dispersion and future market returns across all empirical specifications. The slope coefficients are statistically significant at conventional confidence levels across all estimation procedures and forecast horizons, with the most significant at the 1\% level. Thus, a negative relationship is robust to the exclusion of NBER recessions, the usage of non-overlapping periods, IVX inference, and out-of-sample estimation. 

The predictive ability of kurtosis dispersion is comparable to, or even slightly stronger than, that of skewness dispersion for next month's market returns, but it becomes weaker at longer horizons. Focusing on one-month predictive regressions, $R^2$ values in Table \ref{table: univariate kurtosis} range between 4\% and 6\%, whereas the corresponding values in Table \ref{table: in-sample univariate} tend to be on the lower end of this interval. For longer horizons, the explanatory power of kurtosis dispersion becomes noticeably weaker relative to skewness dispersion, as $R^2$ statistics are smaller by at least two percentage points. In unreported results, we document that using a broader range to compute the cross-sectional kurtosis range further weakens the results. Intuitively, kurtosis is more sensitive to outliers than skewness; hence, extreme kurtosis values tend to be spurious and noisy, providing less valuable information for predicting stock market returns.

\section{Conclusion}
\label{section: conclusion}

This paper proposes a novel predictor of stock market returns, termed skewness dispersion, which measures the inter-percentile range of the cross-sectional distribution of daily realized skewness computed from intraday returns. Elevated values of skewness dispersion indicate a higher degree of heterogeneity in the cross-sectional distribution of return asymmetry.  We demonstrate that the stock market earns lower average returns in the future following the periods of widening skewness dispersion. This negative relationship is robust across estimation procedures and statistical inferences and does not concentrate in recessionary periods. 

The predictive ability of skewness dispersion to forecast future stock market returns is not subsumed by any of the 50 existing predictors. In contrast, only a few alternatives provide incremental information to the cross-sectional range of realized skewness. The out-of-sample forecasts of skewness dispersion also yield substantial economic value for mean-variance investors by producing sizable CER gains and improved Sharpe ratios. Replicating the analysis using kurtosis dispersion, we demonstrate that it also exhibits strong predictive ability, though weaker than skewness dispersion. This highlights a primary role of asymmetry rather than overall tail risk in the return distribution. 

To better understand the economic forces underlying this predictability, we explore several mechanisms. We show that skewness dispersion is related to variation in aggregate risk and investor expectations. Further, its predictive power is concentrated in months with monetary policy decisions, suggesting an important role for the arrival of information. These findings are consistent with an interpretation in which skewness dispersion reflects heterogeneity in beliefs, which is gradually incorporated into prices over time. 

\vspace{20pt}

\begingroup
\linespread{1}
\setlength{\bibsep}{0pt}
\setlength{\bibhang}{1.0em}
\bibliographystyle{chicago}
\bibliography{BIBLIOGRAPHY}
\endgroup

\newpage
\setcounter{section}{0}
\setcounter{equation}{0}
\setcounter{figure}{0}
\setcounter{table}{0}

\def\thesection{\Alph{section}}
\def\thesubsection{\thesection.\arabic{subsection}}
\def\thesubsubsection{\thesubsection.\arabic{subsubsection}}
\renewcommand{\theequation}{\Alph{section}.\arabic{equation}}
\renewcommand{\thetable}{A\arabic{table}}
\renewcommand{\thefigure}{A\arabic{figure}}

\begin{center}
	\Large \textbf{Appendix for} 
\end{center}
\begin{center}
	\Large
	``Skewness Dispersion and Stock Market Returns''
\end{center}

\vspace{10pt}



\section{Supplementary materials}

This appendix presents supplementary details not included in the main body of the paper. Table \ref{tab:variables} summarizes the alternative predictors used in the empirical analysis. Table \ref{table: correlations 1} reports the correlations of skewness dispersion with other predictors. Table \ref{table: in-sample bivariate skewness and cross-section ivx} shows the bivariate predictive regression outputs using the IVX estimation when skewness dispersion is added jointly with other predictors, one at a time.

\definecolor{refblue}{rgb}{0, 0, 1}

\begin{table}[t!]
\caption{Description of alternative predictors}
\begin{minipage}{\textwidth} 
This table describes alternative predictors. For each variable, we reference a source article, a mnemonic, a short name, and a sample period used in our paper. We divide all predictors into five categories: skewness, stock cross-section, sentiment, variance, other macroeconomic and financial variables.
\medskip
\end{minipage}
\renewcommand{\arraystretch}{1.2}
\resizebox{\textwidth}{!}{
\begin{tabular}{rl}
\toprule
\multicolumn{2}{c}{Panel A: Skewness} \\
\midrule
1 & Jondeau, Zhang, Zhu (JFE 2019), \textit{Average skewness matters} \\
   & \textcolor{refblue}{Skm \hspace{1em} market skewness based on daily returns (2000:12 -- 2016:12)} \\
\midrule
2-3 & Jondeau, Zhang, Zhu (JFE 2019), \textit{Average skewness matters} \\
   & \textcolor{refblue}{Skvw, Skew \hspace{1em} average stock value-weighted and equal-weighted skewness (2000:12 -- 2016:12)} \\
\midrule
4 & Andreou, Kagkadis, Philip, Taamouti (JBF 2019), \textit{The information content of forward moments} \\
& \textcolor{refblue}{FSF \hspace{1em} forward skewness factor (2000:12 -- 2015:07)} \\
\midrule
\multicolumn{2}{c}{Panel B: Stock cross-section} \\
\midrule
5 & Andreou, Kagkadis, Maio, Philip (CFR 2021), \textit{Dispersion in options investors' versus analysts'}\\ 
& \textit{expectations: Predictive inference for stock returns} \\
& \textcolor{refblue}{DISP \hspace{1em} dispersion in analyst forecast (2000:12 -- 2022:12)} \\
\midrule
6 & Yu (JFE 2011), \textit{Disagreement and return predictability of stock portfolios} \\
   & \textcolor{refblue}{disag \hspace{1em} analyst forecast disagreements (2000:12 -- 2022:12)} \\
\midrule
7 & Maio (JFM 2016), \textit{Cross-sectional return dispersion and the equity premium} \\
   & \textcolor{refblue}{rdsp \hspace{1em} stock-return dispersion (2000:12 -- 2022:12)} \\
\midrule
8 & Pollett and Wilson (JFE 2010), \textit{Average correlation and stock market returns} \\
   & \textcolor{refblue}{avgcor \hspace{1em} average correlation of daily stock returns (2000:12 -- 2022:12)} \\
\midrule
9 & Kelly and Jiang (RFS 2014), \textit{Tail risk and asset prices} \\
   & \textcolor{refblue}{tail \hspace{1em} tail risk from cross-section (2000:12 -- 2022:12)} \\
\midrule
10 & Kelly and Pruitt (JF 2013), \textit{Market expectations in the cross-section of present values} \\
   & \textcolor{refblue}{fbm \hspace{1em} single factor from B/M cross-section (2000:12 -- 2022:12)} \\
\bottomrule
\end{tabular}%
}
\label{tab:variables}
\end{table}

\definecolor{refblue}{rgb}{0, 0, 1}

\begin{table}[t!]
\caption*{Table \ref{tab:variables} \normalfont -- \textit{Continued}}
\renewcommand{\arraystretch}{1.2}
\resizebox{\textwidth}{!}{
\begin{tabular}{rl}
\toprule
\multicolumn{2}{c}{Panel C: Sentiment} \\
\midrule
11 & Jiang, Lee, Martin, Zhou (JFE 2019), \textit{Manager sentiment and stock returns} \\
& \textcolor{refblue}{MSI \hspace{1em} manager sentiment index (2003:01 -- 2017:12)} \\
\midrule
12-13 & Rapach, Ringgenberg, Zhou (JFE 2016), \textit{Short interest and aggregate stock returns} \\
& \textcolor{refblue}{SII$_{\text{is}}$, SII$_{\text{oos}}$ \hspace{1em} short interest index (2000:12 -- 2022:12)} \\
\midrule
14-15 & Huang, Jiang, Tu, Zhou (RFS 2015), \textit{Investor sentiment aligned: a powerful predictor of stock returns} \\
& \textcolor{refblue}{PLS, PLS$_{\text{orth}}$ \hspace{1em} (orthogonalized) investor sentiment index (2000:12 -- 2022:12)} \\
\midrule
16  & Chen, Tang, Yao, Zhou (JFQA 2019), \textit{Investor attention and stock returns} \\
& \textcolor{refblue}{AI \hspace{1em} attention index (2000:12 -- 2017:12)} \\
\midrule
17 & Han, Li (MS 2021), \textit{Information content of aggregate implied volatility spread} \\
  & \textcolor{refblue}{IVS \hspace{1em} implied volatility spread (2000:12 -- 2015:12)} \\
\midrule
18 & Huang, Jiang, Tu, Zhou (RFS 2015), \textit{Investor sentiment aligned: a powerful predictor of stock returns} \\
   & \textcolor{refblue}{sntm \hspace{1em} optimized investor sentiment index (1965:07 -- 2022:06)} \\
\midrule
\multicolumn{2}{c}{Panel D: Variance} \\
\midrule
19 & Jondeau, Zhang, Zhu (JFE 2019), \textit{Average skewness matters} \\
   & \textcolor{refblue}{Vm \hspace{1em} market variance based on daily returns (2000:12 -- 2016:12)} \\
\midrule
20-21 & Jondeau, Zhang, Zhu (JFE 2019), \textit{Average skewness matters} \\
   & \textcolor{refblue}{Vvw, Vew \hspace{1em} average stock value-weighted and equal-weighted variance (2000:12 -- 2016:12)} \\
\midrule
22 & Andreou, Kagkadis, Philip, Taamouti (JBF 2019), \textit{The information content of forward moments} \\
& \textcolor{refblue}{FVF \hspace{1em} forward variance factor (2000:12 -- 2015:07)} \\
\midrule
23 & Bakshi, Panayotov, Skoulakis (JFE 2011), \textit{Improving the predictability of real economic activity}\\ 
& \textit{and asset returns with forward variances inferred from option portfolios} \\
  & \textcolor{refblue}{impvar \hspace{1em} forward implied variances (2000:12 -- 2022:12)} \\
\midrule
24 & Mrtn (QJE 2017), \textit{Expected Return on the market} \\
   & \textcolor{refblue}{rsvix \hspace{1em} scaled risk-neutral vix (2000:12 -- 2022:12)} \\
\midrule
25 & Bekaert, Hoerova (JE 2021), \textit{The VIX, the variance premium and stock market volatility} \\
  & \textcolor{refblue}{$vp$ \hspace{1em} The VIX squared minus the implied volatility (2000:12 -- 2020:12)} \\
\midrule
26 & Kilic, Shaliastovich (MS 2019), \textit{Good and bad variance premia and expected returns} \\
& \textcolor{refblue}{$vp_{bad}$ \hspace{1em} bad variance risk premium (2000:12 -- 2020:12)} \\
\midrule
27-28 & Babiak, Barunik, Ellington, Bevilacqua (2026), \textit{The common factor in volatility risk premia} \textit{} \\
& \textcolor{refblue}{$cvp$, $cvp_{bad}$ \hspace{1em} (bad) common volatility risk premium  (2000:12 -- 2020:12)} \\
\bottomrule
\end{tabular}%
}
\end{table}

\definecolor{refblue}{rgb}{0, 0, 1}

\begin{table}[t!]
\caption*{Table \ref{tab:variables} \normalfont -- \textit{Continued}}
\renewcommand{\arraystretch}{1.2}
\resizebox{\textwidth}{!}{
\begin{tabular}{rl}
\toprule
\multicolumn{2}{c}{Panel E: Other macroeconomic and financial variables} \\
\midrule
29-30 & Welch, Goyal (RFS 2008), \textit{A comprehensive look at the empirical performance of equity premium prediction} \\ 
& \textcolor{refblue}{dp, dy \hspace{1em} dividend price ratio, dividend yield (2000:12 -- 2022:12)} \\
\midrule
31-32 & Welch, Goyal (RFS 2008), \textit{A comprehensive look at the empirical performance of equity premium prediction} \\  
& \textcolor{refblue}{ep, de \hspace{1em} earnings price ratio, dividend payout ratio (2000:12 -- 2022:12)} \\
\midrule
33 & Welch, Goyal (RFS 2008), \textit{A comprehensive look at the empirical performance of equity premium prediction} \\  
& \textcolor{refblue}{svar \hspace{1em} stock variance (2000:12 -- 2022:12)} \\
\midrule
34 & Welch, Goyal (RFS 2008), \textit{A comprehensive look at the empirical performance of equity premium prediction} \\  
& \textcolor{refblue}{bm \hspace{1em} book to market ratio (2000:12 -- 2022:12)} \\
\midrule
35 & Welch, Goyal (RFS 2008), \textit{A comprehensive look at the empirical performance of equity premium prediction} \\  
& \textcolor{refblue}{ntis \hspace{1em} net equity expansion (2000:12 -- 2022:12)} \\
\midrule
36 & Welch, Goyal (RFS 2008), \textit{A comprehensive look at the empirical performance of equity premium prediction} \\  
& \textcolor{refblue}{tbl \hspace{1em} T-bill rate (2000:12 -- 2022:12)} \\
\midrule
37-38 & Welch, Goyal (RFS 2008), \textit{A comprehensive look at the empirical performance of equity premium prediction} \\  
& \textcolor{refblue}{ltr, lty \hspace{1em} long-term rate (yield) (2000:12 -- 2022:12)} \\
\midrule
39 & Welch, Goyal (RFS 2008), \textit{A comprehensive look at the empirical performance of equity premium prediction} \\  
& \textcolor{refblue}{tms \hspace{1em} term spread (2000:12 -- 2022:12)} \\
\midrule
40 & Welch, Goyal (RFS 2008), \textit{A comprehensive look at the empirical performance of equity premium prediction} \\  
& \textcolor{refblue}{dfy \hspace{1em} default yield spread (2000:12 -- 2022:12)} \\
\midrule
41 & Welch, Goyal (RFS 2008), \textit{A comprehensive look at the empirical performance of equity premium prediction} \\  
& \textcolor{refblue}{dfr \hspace{1em} default return spread (2000:12 -- 2022:12)} \\
\midrule
42 & Welch, Goyal (RFS 2008), \textit{A comprehensive look at the empirical performance of equity premium prediction} \\  
& \textcolor{refblue}{infl \hspace{1em} inflation rate (2000:12 -- 2022:12)} \\
\midrule
43 & Cooper and Priestley (RFS 2009), \textit{Time-varying risk premiums and the output gap} \\
  & \textcolor{refblue}{ogap \hspace{1em} output gap of industrial production (2000:12 -- 2022:12)} \\
\midrule
44 & Jones and Tuzel (RFS 2013), \textit{New orders and asset prices} \\
   & \textcolor{refblue}{ndrbl \hspace{1em} new orders to shipments of durable goods (2000:12 -- 2022:12)} \\
\midrule
45 & Neely, Rapach, Tu, Zhou (MS 2014), \textit{Forecasting the equity risk premium: the role of technical indicators} \\
   & \textcolor{refblue}{tchi \hspace{1em} 14 technical indicators (2000:12 -- 2022:12)} \\
\midrule
46-47 & Li and Yu (JFE 2012), \textit{Investor attention, psychological anchors, and stock return predictability} \\
   & \textcolor{refblue}{dtoy, dtoat \hspace{1em} nearness to Dow 52-week high (2000:12 -- 2022:12)} \\
\midrule
48 & Driesprong, Jacobsen, Maat (JFE 2008), \textit{Striking oil: Another puzzle?} \\
   & \textcolor{refblue}{wtexas \hspace{1em} oil price changes (2000:12 -- 2022:12)} \\
\midrule
49 & Chen, Eaton, Paye (JFE 2018), \textit{Micro(structure) before macro? The predictive power of aggregate illiquidity} \\
  & \textcolor{refblue}{lzrt \hspace{1em} 9 illiquidity measures (2000:12 -- 2022:12)} \\
\midrule
50 & Maio (RF 2013), \textit{The Fed model and the predictability of stock returns} \\
   & \textcolor{refblue}{ygap \hspace{1em} stock-bond yield gap (2000:12 -- 2022:12)} \\
\bottomrule
\end{tabular}%
}
\end{table}

\begin{table}[t!]
\centering
\caption{Correlations}
\begin{minipage}{\textwidth} 
This table provides correlations of ${\tt SD}^{75-25}$ with alternative predictors. Panels A-E report the results for variables belonging to one of the five groups: skewness, stock cross-section, sentiment, variance, and other macroeconomic and financial variables. $^{*},^{**},$ and $^{***}$ indicate significance at the 10\%, 5\%, and 1\% levels, respectively.
\medskip
\end{minipage}
\renewcommand{\arraystretch}{1}
\resizebox{\textwidth}{!}{ 
\begin{tabular}{llllllllllll}
\toprule
& \multicolumn{4}{c}{Panel A: Skewness} & & \multicolumn{6}{c}{Panel B: Stock cross-section} \\
\cmidrule{2-5}\cmidrule{7-12}
&  ${\tt SD}^{75-25}$   & Skm   & Skvw   & Skew & & ${\tt SD}^{75-25}$ & DISP & disag & rdsp & avgcor & tail \\
\midrule
Skm & 0.00 &  &  &  & DISP & -0.05 &  &  &  &  & \\
Skvw & 0.06 & 0.66$^{***}$ &  &  & disag & -0.08 & 0.01 &  &  &  & \\
Skew & 0.15$^{**}$ & 0.50$^{*}$ & 0.81$^{***}$ &   & rdsp & -0.05 & 0.59$^{***}$ & 0.15$^{**}$ &  &  & \\
FSF & -0.13$^{*}$ & -0.06 & -0.23$^{***}$ & -0.29$^{***}$  & avgcor & -0.28$^{***}$ & 0.32$^{***}$ & 0.04 & 0.13$^{**}$ &  & \\
 &  &  &  &  & tail & 0.00 & -0.35$^{***}$ & 0.13$^{**}$ & -0.24$^{***}$ & -0.13$^{**}$ & \\
 &  &  &  &  & fbm & -0.01 & 0.29$^{***}$ & -0.24$^{***}$ & 0.03 & 0.16$^{**}$ & 0.09\\
\midrule
& \multicolumn{10}{c}{Panel C: Sentiment} \\
\cmidrule{2-11}
 & ${\tt SD}^{75-25}$ & MSI & $SII_{is}$ & $SII_{oos}$ &  & $PLS$ & $PLS_{orth}$ & AI & IVS\\
\midrule
MSI & 0.01 &  &  &  &  &  &  &  & \\
SII$_{\text{is}}$ & 0.27$^{***}$ & 0.21$^{***}$ &  &  &  &  &  &  & \\
SII$_{\text{oos}}$ & 0.32$^{***}$ & 0.07 & 0.73$^{***}$ &  &  &  &  &  & \\
PLS & 0.15$^{**}$ & -0.02 & -0.21$^{***}$ & 0.01 &  &  &  &  & \\
PLS$_{\text{orth}}$ & 0.16$^{**}$ & -0.17$^{**}$ & -0.21$^{***}$ & -0.04 &  & 0.93$^{***}$ &  &  & \\
AI & 0.20$^{***}$ & 0.44$^{***}$ & 0.31$^{***}$ & 0.27$^{***}$ &  & 0.44$^{***}$ & 0.38$^{***}$ &  & \\
IVS & -0.10 & -0.23$^{***}$ & -0.11 & -0.11 &  & -0.16$^{**}$ & -0.12$^{*}$ & -0.24$^{***}$ & \\
sntm & -0.24$^{***}$ & 0.50$^{***}$ & 0.01 & 0.07 &  & -0.41$^{***}$ & -0.52$^{***}$ & 0.06 & -0.05\\
\midrule
& \multicolumn{11}{c}{Panel D: Variance} \\
\cmidrule{2-12}
& ${\tt SD}^{75-25}$ & Vm & Vvw & Vew &  & FVF & impvar & rsvix & $vp$ & $vp_{bad}$ & $cvp$\\
\midrule
Vm & -0.16$^{**}$ &  &  &  &  &  &  &  &  &  & \\
Vvw & -0.02 & 0.90$^{***}$ &  &  &  &  &  &  &  &  & \\
Vew & 0.04 & 0.84$^{***}$ & 0.95$^{***}$ &  &  &  &  &  &  &  & \\
FVF & 0.13$^{*}$ & -0.08 & -0.06 & 0.00 &  &  &  &  &  &  & \\
impvar & -0.18$^{***}$ & 0.81$^{***}$ & 0.79$^{***}$ & 0.75$^{***}$ &  & 0.17$^{**}$ &  &  &  &  & \\
rsvix & -0.26$^{***}$ & 0.88$^{***}$ & 0.81$^{***}$ & 0.77$^{***}$ &  & 0.01 & 0.89$^{***}$ &  &  &  & \\
$vp$ & -0.17$^{**}$ & 0.05 & 0.02 & 0.09 &  & 0.15$^{**}$ & 0.28$^{***}$ & 0.40$^{***}$ &  &  & \\
$vp_{bad}$ & -0.27$^{***}$ & 0.58$^{***}$ & 0.50$^{***}$ & 0.52$^{***}$ &  & 0.06 & 0.69$^{***}$ & 0.85$^{***}$ & 0.79 &  & \\
$cvp$ & 0.16$^{**}$ & -0.10 & -0.19$^{***}$ & -0.06 &  & 0.17$^{**}$ & -0.04 & -0.02 & 0.25$^{***}$ & 0.14$^{**}$ & \\
$cvp_{bad}$ & 0.05 & 0.49$^{***}$ & 0.43$^{***}$ & 0.51$^{***}$ &  & 0.14$^{*}$ & 0.49$^{***}$ & 0.57$^{***}$ & 0.35$^{***}$ & 0.56$^{***}$ & 0.75$^{***}$ \\
\bottomrule
\end{tabular}%
}
\label{table: correlations 1}%
\end{table}%

\begin{landscape}
\begin{table}[t!]
\centering
\caption*{\footnotesize Table \ref{table: correlations 1} \normalfont -- \textit{Continued}}
\renewcommand{\arraystretch}{1}
\resizebox{1.4\textwidth}{!}{ 
\begin{tabular}{lrrrrrrrrrrrrrrrrrrrrrr}
\toprule
& \multicolumn{22}{c}{Panel E: Other macroeconomic and financial predictors} \\
\cmidrule{2-23}
  & ${\tt SD}^{75-5}$ & dp & dy & ep & de & svar & bm & ntis & tbl & lty & ltr & tms & dfy & dfr & infl & ogap & ndrbl & tchi & dtoy & dtoat & wtexas & lzrt\\
\midrule
dp & -0.21$^{***}$ &  &  &  &  &  &  &  &  &  &  &  &  &  &  &  &  &  &  &  &  & \\
dy & -0.13$^{**}$ & 0.97$^{***}$ &  &  &  &  &  &  &  &  &  &  &  &  &  &  &  &  &  &  &  & \\
ep & -0.17$^{***}$ & -0.16$^{**}$ & -0.17$^{***}$ &  &  &  &  &  &  &  &  &  &  &  &  &  &  &  &  &  &  & \\
de & 0.06 & 0.53 & 0.53$^{***}$ & -0.92$^{***}$ &  &  &  &  &  &  &  &  &  &  &  &  &  &  &  &  &  & \\
svar & -0.15$^{**}$ & 0.33$^{***}$ & 0.23$^{***}$ & -0.19$^{***}$ & 0.30$^{***}$ &  &  &  &  &  &  &  &  &  &  &  &  &  &  &  &  & \\
bm & -0.17$^{***}$ & 0.77$^{***}$ & 0.76$^{***}$ & 0.16$^{***}$ & 0.17$^{***}$ & 0.10$^{*}$ &  &  &  &  &  &  &  &  &  &  &  &  &  &  &  & \\
ntis & 0.08 & -0.52$^{***}$ & -0.50$^{***}$ & 0.11$^{*}$ & -0.30$^{***}$ & -0.20$^{***}$ & -0.04 &  &  &  &  &  &  &  &  &  &  &  &  &  &  & \\
tbl & 0.13$^{**}$ & -0.34$^{***}$ & -0.36$^{***}$ & 0.12$^{*}$ & -0.24$^{***}$ & -0.08 & -0.40$^{***}$ & -0.11$^{*}$ &  &  &  &  &  &  &  &  &  &  &  &  &  & \\
lty & 0.32$^{***}$ & -0.19$^{***}$ & -0.22$^{***}$ & -0.08 & 0.00 & -0.01 & -0.06 & 0.23$^{***}$ & 0.56$^{***}$ &  &  &  &  &  &  &  &  &  &  &  &  & \\
ltr & -0.05 & 0.12$^{*}$ & 0.07 & 0.05 & 0.00 & 0.19$^{***}$ & 0.10$^{*}$ & 0.03 & 0.02 & -0.02 &  &  &  &  &  &  &  &  &  &  &  & \\
tms & 0.18$^{***}$ & 0.20$^{***}$ & 0.19$^{***}$ & -0.22$^{***}$ & 0.27$^{***}$ & 0.08 & 0.38$^{***}$ & 0.36$^{***}$ & -0.57$^{***}$ & 0.37$^{***}$ & -0.04 &  &  &  &  &  &  &  &  &  &  & \\
dfy & -0.01 & 0.66$^{***}$ & 0.63$^{***}$ & -0.53$^{***}$ & 0.72$^{***}$ & 0.51$^{***}$ & 0.37$^{***}$ & -0.40$^{***}$ & -0.18$^{***}$ & 0.07 & 0.07 & 0.26$^{***}$ &  &  &  &  &  &  &  &  &  & \\
dfr & 0.26$^{***}$ & -0.07 & 0.06 & -0.20$^{***}$ & 0.14$^{**}$ & -0.35$^{***}$ & -0.04 & 0.03 & -0.04 & -0.01 & -0.46$^{***}$ & 0.04 & 0.09 &  &  &  &  &  &  &  &  & \\
infl & -0.04 & -0.27$^{***}$ & -0.27$^{***}$ & 0.02 & -0.12$^{**}$ & -0.28$^{***}$ & -0.15$^{**}$ & 0.11$^{*}$ & 0.05 & 0.01 & -0.27$^{***}$ & -0.04 & -0.27$^{***}$ & 0.00 &  &  &  &  &  &  &  & \\
ogap & 0.23$^{***}$ & -0.29$^{***}$ & -0.33$^{***}$ & 0.26$^{***}$ & -0.34$^{***}$ & -0.12$^{*}$ & -0.17$^{***}$ & 0.11$^{*}$ & 0.67$^{***}$ & 0.76$^{***}$ & 0.09 & 0.01 & -0.20$^{***}$ & -0.15$^{**}$ & 0.02 &  &  &  &  &  &  & \\
ndrbl & -0.16$^{***}$ & -0.30$^{***}$ & -0.29$^{***}$ & 0.57$^{***}$ & -0.61$^{***}$ & -0.27$^{***}$ & -0.04 & 0.14$^{**}$ & 0.11$^{*}$ & -0.03 & -0.05 & -0.16$^{***}$ & -0.49$^{***}$ & -0.07 & 0.13$^{**}$ & 0.24$^{***}$ &  &  &  &  &  & \\
tchi & 0.04 & -0.11$^{*}$ & -0.03 & 0.31$^{***}$ & -0.31$^{***}$ & -0.45$^{***}$ & 0.10 & 0.14$^{**}$ & -0.12$^{*}$ & -0.27$^{***}$ & -0.09 & -0.14$^{*}$ & -0.52$^{***}$ & 0.09 & 0.08 & -0.12$^{*}$ & 0.27$^{***}$ &  &  &  &  & \\
dtoy & 0.13$^{**}$ & -0.45$^{***}$ & -0.36$^{***}$ & 0.62$^{***}$ & -0.71$^{***}$ & -0.58$^{***}$ & -0.14$^{**}$ & 0.19$^{***}$ & 0.05 & -0.19$^{***}$ & -0.13$^{**}$ & -0.24$^{***}$ & -0.77$^{***}$ & 0.11$^{*}$ & 0.13$^{***}$ & 0.10 & 0.47$^{***}$ & 0.71$^{***}$ &  &  &  & \\
dtoat & -0.01 & -0.40$^{***}$ & -0.34$^{***}$ & 0.51$^{***}$ & -0.60$^{***}$ & -0.44$^{***}$ & -0.33$^{***}$ & -0.10$^{*}$ & 0.16$^{***}$ & -0.37$^{***}$ & -0.11$^{*}$ & -0.55$^{***}$ & -0.69$^{***}$ & 0.04 & 0.13$^{**}$ & 0.09 & 0.47$^{***}$ & 0.53$^{***}$ & 0.81$^{***}$ &  &  & \\
wtexas & 0.08 & -0.11$^{*}$ & -0.03 & -0.10 & 0.04 & -0.38$^{***}$ & -0.04 & 0.04 & -0.03 & -0.02 & -0.28$^{***}$ & 0.02 & -0.05 & 0.32$^{***}$ & 0.30$^{***}$ & -0.11$^{*}$ & 0.07 & 0.10 & 0.12$^{**}$ & 0.07 &  & \\
lzrt & 0.07 & -0.29$^{***}$ & -0.24 & 0.06 & -0.17$^{***}$ & -0.42$^{***}$ & -0.24$^{***}$ & 0.11$^{*}$ & -0.10$^{*}$ & -0.24$^{***}$ & -0.25$^{***}$ & -0.12$^{**}$ & -0.37$^{***}$ & 0.27$^{***}$ & 0.16$^{***}$ & -0.26$^{***}$ & 0.06 & 0.25$^{***}$ & 0.30$^{***}$ & 0.29$^{***}$ & 0.09 & \\
ygap & -0.17$^{***}$ & -0.15$^{**}$ & -0.16$^{***}$ & 1.00$^{***}$ & -0.92$^{***}$ & -0.19$^{***}$ & 0.17$^{***}$ & 0.10$^{*}$ & 0.10$^{*}$ & -0.11$^{*}$ & 0.05 & -0.23$^{***}$ & -0.53$^{***}$ & -0.20$^{***}$ & 0.01 & 0.24$^{***}$ & 0.57$^{***}$ & 0.32$^{***}$ & 0.63$^{***}$ & 0.52$^{***}$ & -0.10 & 0.07\\
\bottomrule
\end{tabular}%
}
\end{table}%

\end{landscape}

\begin{table}[t!]
\centering
\caption{In-sample bivariate regression results: IVX estimation} 
\begin{minipage}{\textwidth} 
\footnotesize This table reports the in-sample results of bivariate predictive regressions. The dependent variable is the average monthly market excess returns in logarithm over the $h$-quarter horizon. The independent variables are the skewness dispersion measure $\RSSevenFive$ combined with control variables, one at a time. Panel A, B, C, D, and E present results for five categories: skewness, stock cross-section, sentiment, variance-related, and other macroeconomic and stock market variables from \cite{welch2008comprehensive} and \cite{goyal2024comprehensive}. 
Each panel reports the slope coefficients and IVX-Wald statistics from the \cite{kostakis2015robust} estimation based on the full sample. The 10\%, 5\%, and 1\% critical values for the Wald statistic are 2.706, 3.842, and 6.634. $^{*},^{**},$ and $^{***}$ indicate significance at the 10\%, 5\%, and 1\% levels, respectively.
\medskip
\end{minipage}
\footnotesize
\renewcommand{\arraystretch}{1}
\resizebox{1.0\textwidth}{!}{
\begin{tabular}{lcccccclccccc}
\toprule
&       & $h = 1$     & $h = 3$     & $h = 6$     & $h = 12$    &       &       &       & $h = 1$     & $h = 3$     & $h = 6$     & $h = 12$ \\
\midrule
\multicolumn{13}{l}{Panel A: Skewness} \\
\midrule
$\RSSevenFive$ &       &  -0.122*** &  -0.073*** &  -0.051*** &  -0.028*** &       & $\RSSevenFive$ &       &  -0.119*** &  -0.065*** &  -0.052*** &  -0.038** \\
  &       & [  12.34] & [  18.94] & [  22.14] & [  12.46] &       &   &       & [  11.54] & [  13.93] & [  13.90] & [   5.23] \\
Skm   &       &   0.078 &  -0.365 &   0.003 &   0.022 &       & Skew  &       &  -0.075 &  -0.094 &   0.008 &   0.048 \\
  &       & [   0.02] & [   1.40] & [   0.00] & [   0.02] &       &   &       & [   0.39] & [   1.43] & [   0.01] & [   0.51] \\
[1.0em]
$\RSSevenFive$ &       &  -0.120*** &  -0.068*** &  -0.049*** &  -0.031*** &       & $\RSSevenFive$ &       &  -0.121*** &  -0.060*** &  -0.026** &  -0.019** \\
  &       & [  12.02] & [  16.68] & [  19.51] & [  12.20] &       &   &       & [  10.81] & [  11.88] & [   4.02] & [   6.25] \\
Skvw  &       &  -0.076 &  -0.092** &  -0.019 &   0.018 &       & FSF   &       &   0.055*** &   0.036*** &   0.041*** &   0.034** \\
  &       & [   1.30] & [   5.95] & [   0.46] & [   0.72] &       &   &       & [   6.99] & [  11.19] & [  11.96] & [   6.00] \\
\midrule
\multicolumn{13}{l}{Panel B: Stock cross-section} \\
\midrule
$\RSSevenFive$ &       &  -0.100*** &  -0.059*** &  -0.040*** &  -0.024*** &       & $\RSSevenFive$ &       &  -0.097*** &  -0.050*** &  -0.032*** &  -0.021*** \\
  &       & [  11.39] & [  17.36] & [  17.96] & [  10.46] &       &   &       & [   9.92] & [  11.33] & [  11.79] & [   9.39] \\
DISP  &       &  -0.215 &  -0.119 &  -0.051 &  -0.018 &       & avgcor &       &   0.003 &   0.008 &   0.010 &   0.010** \\
  &       & [   1.23] & [   2.33] & [   1.16] & [   0.47] &       &   &       & [   0.01] & [   0.64] & [   2.03] & [   4.90] \\
[1.0em]
$\RSSevenFive$ &       &  -0.097*** &  -0.057*** &  -0.041*** &  -0.031*** &       & $\RSSevenFive$ &       &  -0.099*** &  -0.057*** &  -0.040*** &  -0.028*** \\
  &       & [  10.71] & [  16.10] & [  17.65] & [  13.62] &       &   &       & [  11.14] & [  16.57] & [  16.85] & [  10.53] \\
disag &       &   0.003 &  -0.000 &  -0.001 &  -0.001 &       & tail  &       &   0.067 &  -0.011 &  -0.005 &  -0.007 \\
  &       & [   0.69] & [   0.01] & [   0.64] & [   1.98] &       &   &       & [   0.38] & [   0.06] & [   0.04] & [   0.19] \\
[1.0em]
$\RSSevenFive$ &       &  -0.099*** &  -0.057*** &  -0.039*** &  -0.026*** &       & $\RSSevenFive$ &       &  -0.099*** &  -0.057*** &  -0.038*** &  -0.024*** \\
  &       & [  11.03] & [  16.57] & [  17.56] & [  12.96] &       &   &       & [  11.39] & [  16.93] & [  16.82] & [   8.50] \\
rdsp  &       &  -0.107 &  -0.071 &  -0.019 &  -0.002 &       & fbm   &       &  -0.081** &  -0.029** &  -0.016* &  -0.003 \\
  &       & [   0.19] & [   0.48] & [   0.12] & [   0.01] &       &   &       & [   3.85] & [   3.85] & [   3.22] & [   0.24] \\
\bottomrule
\end{tabular}%
}
\label{table: in-sample bivariate skewness and cross-section ivx}%
\end{table}%

\begin{table}[t!]
\centering
\caption*{\footnotesize Table \ref{table: in-sample bivariate skewness and cross-section ivx} \normalfont -- \textit{Continued}}
\footnotesize
\renewcommand{\arraystretch}{1}
\resizebox{1.0\textwidth}{!}{
\begin{tabular}{lcccccclccccc}
\toprule
&       & $h = 1$     & $h = 3$     & $h = 6$     & $h = 12$    &       &       &       & $h = 1$     & $h = 3$     & $h = 6$     & $h = 12$ \\
\midrule
\multicolumn{13}{l}{Panel C: Sentiment} \\
\midrule
$\RSSevenFive$ &       &  -0.100*** &  -0.058*** &  -0.047*** &  -0.023** &       & $\RSSevenFive$ &       &  -0.087*** &  -0.050*** &  -0.033*** &  -0.021*** \\
  &       & [   9.28] & [  12.39] & [  17.56] & [   4.67] &       &   &       & [   8.66] & [  12.29] & [  12.02] & [   6.88] \\
MSI   &       &  -0.008** &  -0.003*** &  -0.002*** &  -0.001** &       & PLS$_{\text{orth}}$ &       &  -0.007 &  -0.002 &  -0.001 &  -0.001 \\
  &       & [   6.54] & [   6.89] & [   6.67] & [   5.75] &       &   &       & [   2.06] & [   2.02] & [   2.13] & [   1.26] \\
[1.0em]
$\RSSevenFive$ &       &  -0.094*** &  -0.056*** &  -0.040*** &  -0.028*** &       & $\RSSevenFive$ &       &  -0.097*** &  -0.049*** &  -0.035*** &  -0.020*** \\
  &       & [   9.43] & [  13.70] & [  14.11] & [  11.36] &       &   &       & [   8.77] & [   9.37] & [  10.98] & [   7.20] \\
SII$_{\text{is}}$ &       &  -0.002 &  -0.000 &   0.000 &   0.000 &       & AI &       &  -0.049*** &  -0.021*** &  -0.012*** &  -0.007*** \\
  &       & [   0.52] & [   0.11] & [   0.02] & [   0.51] &       &   &       & [   8.82] & [  10.10] & [   8.68] & [   9.06] \\
[1.0em]
$\RSSevenFive$ &       &  -0.082*** &  -0.048*** &  -0.034*** &  -0.026*** &       & $\RSSevenFive$ &       &  -0.126*** &  -0.064*** &  -0.042*** &  -0.028*** \\
  &       & [   7.04] & [   9.75] & [   9.55] & [   6.76] &       &   &       & [  12.91] & [  14.73] & [  14.17] & [  12.11] \\
SII$_{\text{oos}}$ &       &  -0.005* &  -0.001 &  -0.000 &   0.000 &       & IVS   &       &   1.101*** &   0.654*** &   0.404*** &   0.087 \\
  &       & [   3.80] & [   2.56] & [   1.04] & [   0.00] &       &   &       & [  14.87] & [  22.39] & [  15.79] & [   1.24] \\
[1.0em]
$\RSSevenFive$ &       &  -0.090*** &  -0.050*** &  -0.033*** &  -0.020** &       & $\RSSevenFive$ &       &  -0.102*** &  -0.064*** &  -0.054*** &  -0.031*** \\
  &       & [   9.15] & [  12.56] & [  12.18] & [   6.41] &       &   &       & [  11.61] & [  18.94] & [  25.75] & [  17.58] \\
PLS   &       &  -0.005 &  -0.002 &  -0.001 &  -0.001 &       & sntm  &       &  -0.009 &  -0.004 &  -0.003 &  -0.002 \\
  &       & [   1.21] & [   1.74] & [   2.35] & [   2.68] &       &   &       & [   0.85] & [   1.53] & [   2.07] & [   2.24] \\
\midrule
\multicolumn{13}{l}{Panel D: Variance-related} \\
\midrule
$\RSSevenFive$ &       &  -0.142*** &  -0.076*** &  -0.049*** &  -0.029*** &       & $\RSSevenFive$ &       &  -0.104*** &  -0.058*** &  -0.039*** &  -0.030*** \\
  &       & [  16.95] & [  21.58] & [  22.07] & [  14.24] &       &   &       & [  11.22] & [  16.06] & [  17.19] & [  15.38] \\
Vm    &       &  -0.377*** &  -0.129*** &  -0.036 &   0.015 &       & rsvix &       &  -0.054 &  -0.011 &   0.011 &   0.019* \\
  &       & [  10.57] & [   7.85] & [   1.65] & [   0.50] &       &   &       & [   0.53] & [   0.16] & [   0.44] & [   3.34] \\
[1.0em]
$\RSSevenFive$ &       &  -0.124*** &  -0.067*** &  -0.045*** &  -0.027*** &       & $\RSSevenFive$ &       &  -0.106*** &  -0.067*** &  -0.048*** &  -0.030*** \\
  &       & [  13.01] & [  16.60] & [  15.45] & [   6.96] &       &   &       & [  10.86] & [  19.39] & [  23.00] & [  13.85] \\
Vvw   &       &  -0.676 &  -0.436* &  -0.142 &  -0.011 &       & vp &       &   0.004 &   0.011* &   0.009** &   0.006** \\
  &       & [   1.65] & [   3.22] & [   0.97] & [   0.01] &       &   &       & [   0.13] & [   3.52] & [   6.52] & [   5.44] \\
[1.0em]
$\RSSevenFive$ &       &  -0.120*** &  -0.066*** &  -0.048*** &  -0.029*** &       & $\RSSevenFive$ &       &  -0.112*** &  -0.067*** &  -0.048*** &  -0.030*** \\
  &       & [  12.11] & [  15.77] & [  17.35] & [   8.31] &       &   &       & [  11.54] & [  19.04] & [  22.92] & [  12.74] \\
Vew   &       &  -0.319 &  -0.176 &  -0.035 &   0.009 &       & vp$_{\text{bad}}$ &       &  -0.007 &   0.004 &   0.006* &   0.006** \\
  &       & [   1.04] & [   1.70] & [   0.21] & [   0.03] &       &   &       & [   0.19] & [   0.44] & [   2.95] & [   3.99] \\
[1.0em]
$\RSSevenFive$ &       &  -0.144*** &  -0.076*** &  -0.061*** &  -0.039*** &       & $\RSSevenFive$ &       &  -0.128*** &  -0.066*** &  -0.036*** &  -0.021*** \\
  &       & [  14.84] & [  18.61] & [  25.93] & [  13.98] &       &   &       & [  16.74] & [  20.49] & [  14.70] & [   9.23] \\
FVF   &       &   0.043** &   0.019** &   0.014** &   0.007* &       & cvp   &       &   0.271*** &   0.112*** &   0.097*** &   0.078*** \\
  &       & [   4.42] & [   5.21] & [   6.45] & [   3.40] &       &   &       & [  11.49] & [   7.30] & [  10.49] & [  12.71] \\
[1.0em]
$\RSSevenFive$ &       &  -0.104*** &  -0.059*** &  -0.039*** &  -0.028*** &       & $\RSSevenFive$ &       &  -0.111*** &  -0.066*** &  -0.046*** &  -0.030*** \\
  &       & [  11.25] & [  16.73] & [  17.24] & [  12.80] &       &   &       & [  12.51] & [  19.83] & [  21.56] & [  13.82] \\
impvar &       &  -0.813 &  -0.386 &  -0.036 &   0.080 &       & cvp$_{\text{bad}}$ &       &   0.146* &   0.039 &   0.041** &   0.042*** \\
  &       & [   0.64] & [   0.99] & [   0.03] & [   0.31] &       &   &       & [   3.57] & [   1.60] & [   4.86] & [   8.17] \\
\bottomrule
\end{tabular}%
}
\label{table: in-sample bivariate sentiment and variance ivx}%
\end{table}%

\begin{table}[t!]
\centering
\caption*{\footnotesize Table \ref{table: in-sample bivariate skewness and cross-section ivx} \normalfont -- \textit{Continued}}
\footnotesize
\renewcommand{\arraystretch}{1}
\resizebox{1.0\textwidth}{!}{
\begin{tabular}{lcccccclccccc}
\toprule
&       & $h = 1$     & $h = 3$     & $h = 6$     & $h = 12$    &       &       &       & $h = 1$     & $h = 3$     & $h = 6$     & $h = 12$ \\
\midrule
\multicolumn{13}{l}{Panel E: Other macroeconomic and financial variables} \\
\midrule
$\RSSevenFive$ &       &  -0.089*** &  -0.050*** &  -0.033*** &  -0.019*** &       & $\RSSevenFive$ &       &  -0.093*** &  -0.053*** &  -0.036*** &  -0.023*** \\
 &       & [   8.65] & [  12.61] & [  12.73] & [   7.00] &       &  &       & [   9.84] & [  14.42] & [  15.06] & [   9.64] \\
dp    &       &   0.020 &   0.009 &   0.007 &   0.004 &       & tbl   &       &  -0.328 &  -0.116 &  -0.068 &  -0.045** \\
 &       & [   0.59] & [   1.05] & [   2.07] & [   2.59] &       &  &       & [   1.73] & [   1.92] & [   2.47] & [   4.04] \\
[1.0em]
$\RSSevenFive$ &       &  -0.090*** &  -0.049*** &  -0.032*** &  -0.018** &       & $\RSSevenFive$ &       &  -0.077** &  -0.047*** &  -0.035*** &  -0.030** \\
 &       & [   9.38] & [  12.10] & [  11.91] & [   5.98] &       &  &       & [   6.22] & [   9.13] & [   9.55] & [   6.14] \\
dy    &       &   0.029 &   0.011 &   0.007 &   0.004* &       & lty   &       &  -0.555** &  -0.134 &  -0.036 &   0.018 \\
 &       & [   1.29] & [   1.61] & [   2.63] & [   2.79] &       &  &       & [   5.07] & [   2.35] & [   0.55] & [   0.25] \\
[1.0em]
$\RSSevenFive$ &       &  -0.099*** &  -0.059*** &  -0.042*** &  -0.030*** &       & $\RSSevenFive$ &       &  -0.097*** &  -0.057*** &  -0.041*** &  -0.026*** \\
 &       & [  10.80] & [  17.22] & [  19.10] & [  14.54] &       &  &       & [  10.63] & [  16.55] & [  19.80] & [  13.74] \\
ep    &       &  -0.001 &  -0.002 &  -0.002 &  -0.001 &       & ltr   &       &   0.060 &   0.007 &   0.038 &   0.028 \\
 &       & [   0.01] & [   0.61] & [   1.59] & [   1.73] &       &  &       & [   0.49] & [   0.01] & [   1.03] & [   0.52] \\
[1.0em]
$\RSSevenFive$ &       &  -0.099*** &  -0.058*** &  -0.041*** &  -0.030*** &       & $\RSSevenFive$ &       &  -0.097*** &  -0.058*** &  -0.042*** &  -0.034*** \\
 &       & [  11.05] & [  16.84] & [  18.37] & [  14.58] &       &  &       & [  10.41] & [  15.88] & [  17.84] & [  16.16] \\
de    &       &   0.003 &   0.003 &   0.002* &   0.002** &       & tms   &       &  -0.032 &   0.029 &   0.043 &   0.054** \\
 &       & [   0.26] & [   1.52] & [   3.58] & [   4.61] &       &  &       & [   0.02] & [   0.15] & [   1.15] & [   5.31] \\
[1.0em]
$\RSSevenFive$ &       &  -0.102*** &  -0.060*** &  -0.039*** &  -0.029*** &       & $\RSSevenFive$ &       &  -0.099*** &  -0.057*** &  -0.039*** &  -0.027*** \\
 &       & [  11.41] & [  17.60] & [  17.22] & [  15.13] &       &  &       & [  11.11] & [  16.56] & [  16.94] & [  13.09] \\
svar  &       &  -0.332 &  -0.209 &  -0.006 &   0.134 &       & dfy   &       &  -0.606 &  -0.040 &   0.116 &   0.183** \\
 &       & [   0.65] & [   1.30] & [   0.00] & [   2.45] &       &  &       & [   0.82] & [   0.03] & [   0.85] & [   4.46] \\
[1.0em]
$\RSSevenFive$ &       &  -0.087*** &  -0.049*** &  -0.033*** &  -0.020*** &       & $\RSSevenFive$ &       &  -0.110*** &  -0.059*** &  -0.035*** &  -0.023*** \\
 &       & [   8.53] & [  12.33] & [  12.67] & [   8.34] &       &  &       & [  12.97] & [  17.11] & [  14.52] & [  10.98] \\
bm    &       &   0.078 &   0.033* &   0.022** &   0.012** &       & dfr   &       &   0.207 &   0.173** &   0.142*** &   0.098** \\
 &       & [   2.09] & [   3.25] & [   5.29] & [   5.60] &       &  &       & [   2.17] & [   4.23] & [   6.86] & [   4.01] \\
[1.0em]
$\RSSevenFive$ &       &  -0.101*** &  -0.060*** &  -0.042*** &  -0.031*** &       & $\RSSevenFive$ &       &  -0.098*** &  -0.057*** &  -0.042*** &  -0.027*** \\
 &       & [  11.53] & [  17.99] & [  20.32] & [  16.16] &       &  &       & [  10.93] & [  16.49] & [  20.15] & [  14.79] \\
ntis  &       &   0.212 &   0.080 &   0.041 &   0.021 &       & infl  &       &   0.105 &   0.009 &  -0.322 &  -0.488** \\
 &       & [   1.88] & [   2.36] & [   2.24] & [   1.84] &       &  &       & [   0.02] & [   0.00] & [   1.52] & [   4.70] \\
\midrule
$\RSSevenFive$ &       &  -0.086*** &  -0.047*** &  -0.032*** &  -0.022*** &       & $\RSSevenFive$ &       &  -0.097*** &  -0.058*** &  -0.043*** &  -0.039*** \\
 &       & [   8.09] & [  10.41] & [  10.34] & [   7.07] &       &  &       & [  10.61] & [  16.23] & [  18.14] & [  15.90] \\
ogap  &       &  -0.112* &  -0.040** &  -0.018* &  -0.008 &       & dtoat &       &  -0.025 &  -0.013 &  -0.011** &  -0.011*** \\
 &       & [   3.76] & [   4.06] & [   3.08] & [   1.78] &       &  &       & [   0.96] & [   2.17] & [   5.07] & [   9.95] \\
[1.0em]
$\RSSevenFive$ &       &  -0.098*** &  -0.059*** &  -0.044*** &  -0.034*** &       & $\RSSevenFive$ &       &  -0.098*** &  -0.055*** &  -0.036*** &  -0.025*** \\
 &       & [  10.78] & [  17.29] & [  20.74] & [  19.29] &       &  &       & [  10.79] & [  15.67] & [  15.68] & [  13.48] \\
ndrbl &       &  -0.003 &  -0.021 &  -0.022 &  -0.021** &       & wtexas &       &   0.002 &   0.029 &   0.014 &   0.011 \\
 &       & [   0.00] & [   0.69] & [   2.61] & [   5.83] &       &  &       & [   0.00] & [   2.58] & [   0.67] & [   0.55] \\
[1.0em]
$\RSSevenFive$ &       &  -0.101*** &  -0.058*** &  -0.038*** &  -0.028** &       & $\RSSevenFive$ &       &  -0.101*** &  -0.057*** &  -0.040*** &  -0.026*** \\
 &       & [  11.32] & [  16.62] & [  16.28] & [   6.35] &       &  &       & [  11.50] & [  16.76] & [  17.61] & [  13.38] \\
tchi  &       &   0.003 &   0.001 &   0.001 &  -0.000 &       & lzrt  &       &   0.006 &   0.001 &  -0.001 &  -0.001 \\
 &       & [   1.74] & [   1.38] & [   0.77] & [   0.03] &       &  &       & [   0.46] & [   0.03] & [   0.41] & [   1.64] \\
[1.0em]
$\RSSevenFive$ &       &  -0.098*** &  -0.057*** &  -0.039*** &  -0.034*** &       & $\RSSevenFive$ &       &  -0.098*** &  -0.059*** &  -0.042*** &  -0.031*** \\
 &       & [  10.47] & [  15.73] & [  16.77] & [  14.35] &       &  &       & [  10.71] & [  17.14] & [  19.05] & [  14.40] \\
dtoy  &       &  -0.002 &  -0.002 &  -0.006 &  -0.011** &       & ygap  &       &  -0.000 &  -0.002 &  -0.002 &  -0.001 \\
 &       & [   0.00] & [   0.03] & [   0.83] & [   5.09] &       &  &       & [   0.00] & [   0.54] & [   1.51] & [   1.71] \\
\bottomrule
\end{tabular}%
}
\label{table: in-sample bivariate goyal welth ivx}%
\end{table}%

\end{document}